\documentclass[twocolumn,showpacs,preprintnumbers,amsmath,amssymb]{revtex4}

\usepackage{amsmath}
\usepackage{natbib}
\usepackage{graphicx}
\usepackage{epstopdf}
\usepackage{subfigure}
\usepackage{dcolumn}
\usepackage{bm}
\usepackage{color}
\begin{document}

\title{\large Effects of two quantum correction parameters on chaotic dynamics of
particles near renormalized group improved Schwarzschild black holes}

\author{Junjie Lu$^{1,2}$}
\author{Xin Wu$^{1,2}$}
\email{21200006@sues.edu.cn; wuxin_1134@sina.com}
 \affiliation{$^{1}$School of Mathematics, Physics and Statistics, Shanghai
University of Engineering Science, Shanghai 201620, China
\\ $^{2}$Center of Application and Research of Computational Physics,
Shanghai University of Engineering Science, Shanghai 201620, China}

\begin{abstract}

\textbf{Abstract:}A renormalized group improved Schwarzschild black hole spacetime contains two quantum correction parameters. One parameter $\gamma$ represents
the identification of cutoff of the distance scale, and another  parameter $\Omega$ stems
from nonperturbative renormalization group theory.
The two parameters are constrained by the data from the shadow of M87* central black hole.
The dynamics of electrically charged test particles around the black hole are integrable. However,
when the black hole is immersed in an external asymptotically uniform magnetic field, the
dynamics are not integrable and may allow for the occurrence of chaos. 
Employing an explicit symplectic integrator, we survey  the contributions of the two parameters to the chaotic dynamical behavior.
It is found that a small change of the parameter $\gamma$ constrained by the shadow of
M87* black hole has an almost negligible effect on the dynamical transition of
particles from order to chaos. However, a small decrease in the parameter
$\Omega$ leads to an enhancement in the strength of chaos from the global phase space
structure.
A theoretical interpretation is given to the different contributions.
The term with the parameter $\Omega$ dominates the term with the parameter
$\gamma$, even if the two parameters have same values. 
In particular, the parameter $\Omega$  acts as a repulsive force, and its decrease means
a weakening of the repulsive force or equivalently  enhancing the attractive
force from the black hole. On the other hand, there is a positive Lyapunov exponent that
is universally given by the surface gravity of the black hole when
$\Omega\geq 0$ is small and the external magnetic field vanishes.
In this case, the horizon would influence chaotic behavior in the
motion of charged particles around the black hole surrounded by
the external magnetic field. This point can explain why a smaller
value of the renormalization group parameter would much easily
induce chaos than a larger value.


\end{abstract}



\maketitle

\section{Introduction}

The images of the shadows of the supermassive black holes M87* at the
center of the giant elliptical galaxy and Sgr A* in the Galactic
Center~\cite{r1,r2} directly confirm the existence of black holes. They
have not only been successful for testing Einstein's theory of general
relativity in the strong field regime but have also been useful for testing
other theories of gravity. 
Although the theory of general
relativity has been proven to  be a quite effective theory of
gravity, it cannot completely  describe various situations like
the accelerating expansion of the universe and galaxy rotation
curves~\cite{r3}. In this sense, general relativity is not the final
theory of gravity and is necessarily extended to  cure its
shortcomings. Modified theories of gravity can be considered as
generalizations of general relativity; in fact, there have been
a number of modified gravity theories, such as $f$($R$) gravity~\cite{r4},
scalar--tensor theory, tensor--vector--scalar theory,
Ho\v{r}ava--Lifshitz gravity, Gauss--Bonnet theory, Einstein--\AE
ther theory and quantum gravity (see, e.g., review papers~\cite{r5,r6} for
more details on these modified gravity theories).

In addition to the mentioned-above limitations in application, the existence of
curvature singularities is one of the shortcomings of the theory of general relativity.
Because these singularities are hidden behind event horizons,  the pathologies at the singular region
have no causal connection with  the physics in the exterior region. The motion of physical objects outside
the event horizons can be accurately described by general relativity black hole solutions.
However, how physical objects fall inside  is unknown due to  the singular behavior of
general relativity black hole solutions.
Thus, the whole spacetime that contains both the exterior of a black hole and the interior of black hole
is desired. Some regular black hole solutions can avoid the black hole singularity problem.
Bardeen black holes~\cite{r7} are a class  of regular black hole
models that are not exact solutions of the Einstein field equations. Of course, there are also
singularity-free
black hole solutions that are  exact solutions of the Einstein field equations  coupled to
suitable nonlinear electrodynamics~\cite{r8}.
The quantum gravity effects from loop quantum gravity and the renormalization group
improvement (RGI) technique
can also eliminate the emergence of black hole singularities.  Bonanno and Reuter~\cite{r9} gave
an RGI Schwarzschild  black hole obtained from the renormalization group improvement
of
the
Einstein--Hilbert action with a running Newton constant. This black hole solution is based
on
the quantum effects in Schwarzschild black hole geometry. In practice, this solution is
simply the Schwarzschild  black hole, where Newton's constant is modified as a variable
quantity depending on distance or time. 
There are two quantum correction parameters, including one parameter $\gamma$
regarding
the identification of cutoff of the distance scale and another  parameter $\Omega$
stemming from nonperturbative renormalization group theory.
\textcolor{black}{When $\Omega$ is replaced with $-\Omega$, the RGI Schwarzschild black holes become scale-dependent Planck stars~\cite{r10}. }

Taking the lenses of the supermassive black holes Sgr A* and M87*,
the authors of~\cite{r11} investigated weak and strong deflection
gravitational lensing of the RGI Schwarzschild black hole. They
also provided a constraint in the quantum effects on the two
quantum correction parameters  $\gamma$ and $\Omega$. The authors
of~\cite{r12} discussed the circular motion of electrically charged
particles around the RGI Schwarzschild black hole in the presence
of an external asymptotically uniform magnetic field. They found that
the  radius of innermost stable circular orbit (ISCO) increases
with an increase in the parameter $\gamma$ but decreases with
an increase in the parameter $\Omega$ or the magnetic field.
\textcolor{black}{RGI Schwarzschild black holes and the scale-dependent Planck stars
can be distinguished in terms of the particles' periodic motions based on
the precession of the S2 star from the GRAVITY observations, the
shadow result of Sgr A*  from EHT, and the gravitational waveforms~\cite{r10}. }
The authors of~\cite{r13} showed that the variation in the parameters of the
RGI Schwarzschild black hole causes a transition from periodic
motions of time-like particles around the black hole to
quasi-periodic motions. With the aid of  the 3:2 resonance in
X-ray binaries, quasi-periodic oscillation frequencies for the
epicyclic motions of charged particles around the black hole
immersed in an external asymptotically uniform magnetic field were
studied. \textcolor{black}{Such high frequency quasi-periodic oscillations in black hole systems
arise from  resonant phenomena  between orbital
and epicyclic motion of accreting matter~\cite{r14,r15,r16}. }
Periodic orbits and the epicyclic frequencies of
quasi-circular motions around other quantum-corrected black holes
were also considered in Refs.~\cite{r17,r18,r19}.

Although the external asymptotically uniform magnetic field is too small to affect the black hole geometry,
it can exert an important influence on the trajectories of charged particles moving near the black hole.
The trajectories are nonintegrable, and even allow for the chaotic behavior.
Chaos is one kind of nonlinear phenomenon of a dynamical system, which exhibits sensitive dependence on
initial conditions. The existence of
chaos in some general relativity or modified gravity black holes has been reported in the literature, e.g.,~\cite{r20,r21,r22,r23,r24,r25}.
Chaotic behavior may be useful to explain some astrophysical phenomena.
Chaotic scattering of charged particles arises  from the black hole gravitational field combined with the external uniform magnetic
field. It provides a
mechanism for charged particle acceleration along the magnetic field lines and is helpful for the formation of  relativistic jets~\cite{r26}.
In addition, self-similar fractal structures for characterizing the nature of chaotic photon motions can be observed
in the shadow of nonintegrable black~hole~spacetimes~\cite{r27,r28}.

As claimed above, the effect of varying the RGI black hole parameters on the ISCO dynamics of the charged particles was
shown in Ref.~\cite{r12}, and the transition from periodic motions of time-like particles
around the RGI black hole to quasi-periodic motions
was studied in Ref.~\cite{r13}. Unlike the two works,
the present paper mainly considers how varying the RGI black hole parameters causes
a transition from regular dynamics to chaotic dynamics
according to the constraint of the two quantum correction parameters $\gamma$ and $\Omega$ in Ref.~\cite{r11}.
For this purpose, we briefly introduce the dynamical equations of charged particles
around the RGI black hole surrounded by an external asymptotically uniform magnetic field in Section 2.
Then, we numerically investigate the regular and chaotic dynamics of
charged particles in Section 3.  Finally, the main results are summarized in Section 4.

\section{Dynamical model of charged particles}

An RGI Schwarzschild black hole metric~\cite{r9} is briefly introduced.
Then, an external asymptotically  uniform magnetic field surrounding the black hole
is described in terms of electromagnetic four-potential. Dynamical equations of charged particles
around the black hole  immersed in such an external magnetic field are obtained from
a Hamiltonian system.

\subsection{ RGI  Schwarzschild  black hole metric}

Suppose that the Newton's gravitational constant is $G_0$. Bonanno and Reuter~\cite{r9} modified the gravitational constant as
a position-dependent quality
\begin{eqnarray}
 G(r)= \frac{G_0r^3}{r^3+\Omega G_0(r+\gamma G_0M)},
 \end{eqnarray}
where  $M$ is the mass of Schwarzschild
black hole. $\gamma$ and $\Omega$ are two new spatio-temporal quantum correction
parameters without dimensions. The former parameter corresponds to
the identification of cutoff of the distance scale, and the latter  parameter corresponds to
nonperturbative renormalization group theory.
In this case, the function $f(r)=1-2G(r)M/c^2$ is written as
\begin{eqnarray}
       f(r) = 1-\frac{2 M}{r} {\left(\frac{\gamma \Omega  M^3}{r^3}+\frac{M^2 \Omega }{r^2}+1\right)}^{-1},
    \end{eqnarray}
where the original  Newton's gravitational constant $G_0$ and the speed of light $c$ are taken as geometric units,
$G_0=c=1$. On the other hand,
quantum effects are included in the geometry of Schwarzschild
black hole by means of  the techniques of renormalization group.
Considering the two points, Bonanno and Reuter obtained  an RGI Schwarzschild black
hole~metric
 \begin{eqnarray}
 ds^2 &=& g_{\mu \nu}dx^\mu dx^\nu \nonumber \\
  &=& -f(r)\text{dt}^2+ f(r)^{-1}{\text{dr}^2} \nonumber \\
  &&+r^2 \left(d\varphi^2 \sin^2\theta+d\theta^2\right).
 \end{eqnarray}

The two  quantum correction parameters are constrained by the data from the shadow of
the M87 central black hole.
The data from the shadow of the M87 central black hole give the constraints of the two
parameters
in the ranges of $0.02 \leq \gamma \leq 0.22$ and $0.165 \leq \Omega \leq 9.804$~\cite{r11}.

Unlike Schwarzschild spacetime with \textcolor{black}{$\gamma=\Omega=0$},
RGI Schwarzschild spacetime with \textcolor{black}{$\gamma\neq0$ and
$\Omega\neq0$} has finite curvature values of  the Ricci
scalar, square of the Ricci tensor and the Kretschmann scalar  t $r = 0$, and
$\gamma\neq0$~\cite{r12}. That is to say,  the quantum gravity effects with the loop quantum gravity and the RGI
approach completely rule out the formation of the black hole singularity. However,
the existence of one or two  horizons is still possible in the RGI Schwarzschild spacetime.
In fact, there are  two horizons for $0<\varepsilon<1$,  a single horizon for
$\varepsilon=1$, and  no horizon for $\varepsilon>1$,
where $\Omega=\varepsilon\Omega_+$. $\Omega_+$ is a  root of the
equation $f(r)=0$ in the following form
\begin{eqnarray}
\Omega_+ =-\frac{27}{8}\gamma^2-\frac{9}{2}\gamma+\frac{1}{2}+\frac{1}{8}\sqrt{(\gamma+2)(9\gamma+2)^3}.
 \end{eqnarray}

\subsection{Electromagnetic four-potential}

An external asymptotically uniform magnetic field surrounding the RGI
Schwarzschild black hole is assumed to be perpendicular to the equatorial plane $\theta=\pi/2$.
The magnetic field has an asymptotic value $B$.  The spacetime of an RGI Schwarzschild
black hole
is \textcolor{black}{asymptotically} flat, and has two conserved
Killing vectors along the time-like and space-like directions, $\xi^{\mu}_{t}=(1,0,0,0)$ and
$\xi^{\mu}_{\varphi}=(0,0,0,1)$.
According to the Wald method~\cite{r29},  an electromagnetic four-potential is a linear
combination of the two Killing vectors in the form
\begin{eqnarray}
   A^{\mu}=c_t\xi^{\mu}_{t}+c_{\varphi}\xi^{\mu}_{\varphi},
\end{eqnarray}
where $c_t$ and $c_{\varphi}$ are constant coefficients.
$A^{\mu}$ is a solution of the
source-less Maxwell field equations.
Taking  $c_t=0$ and $c_{\varphi}=B/2$,
the authors of~\cite{r12} gave an electromagnetic four-potential with only one nonzero
covariant component
\begin{eqnarray}
    A_\varphi=\frac{B}{2}g_{\mu\varphi}\xi^{\mu}=\frac{B}{2}g_{\varphi\varphi} =\frac{1}{2}Br^2\sin^2\theta.
\end{eqnarray}

This electromagnetic four-potential is used to describe the asymptotically uniform magnetic field.

Because the spacetime of an RGI Schwarschild black hole deals with the quantum gravity
correction,
it belongs to a kind of modified gravity theory. In general, such a modified gravity black hole is a
nonvacuum solution or is non-Ricci-flat. In this case, the two constant coefficients $c_t$
and $c_{\varphi}$ from the Wald's formulas
should be generalized as functions of the coordinates so that the obtained  four-potential
is an exact solution to the source-less Maxwell equations, as was claimed by Azreg-A\"{i}nou~\cite{r30}.
See also the latest paper by Cao et al.~\cite{r31}.
Does the covariant potential (6) not satisfy the source-less Maxwell equations?
To answer this question, we rewrite Equation (2) as
$f(r)=1-(2/r^2)\tilde{f}(r)$, where $\tilde{f}(r)$ is a function
\begin{eqnarray}
       \tilde{f}(r) =rM \left(\frac{\gamma \Omega  M^3}{r^3}+\frac{M^2 \Omega }{r^2}+1\right)^{-1}.
    \end{eqnarray}

Considering $1/r$ as a small value,
we expand the function $\tilde{f}(r)$
as
\begin{eqnarray}
\tilde{f}(r) &\approx& rM\left(1-\frac{M^2 \Omega }{r^2}-\frac{\gamma \Omega  M^3}{r^3}+\cdots \right) \nonumber \\
&=& rM -\frac{M^3 \Omega }{r}-\frac{\gamma \Omega  M^4}{r^2}+\cdots.
\end{eqnarray}

This expression is an asymptotical expansion function with $f_1=M$ and $f_2=0$ mentioned in Ref.~\cite{r30}.
Noticing the neutral black hole with charge $Q=0$ in \textcolor{black}{Equation (23) of~\cite{r30}}, we can easily know that the covariant potential component (6)
is what we want. In other words, the potential $A_\varphi$ satisfies the Maxwell equations.

\subsection{Hamiltonian system}

The magnetic field given by the four-potential (6) is so weak that it has a negligible effect
on the spacetime geometry. In spite of this, it strongly affects the
motion of a charged test particle with mass $m$ and charge $q$.
The motion is governed by the following Hamiltonian
\begin{eqnarray}
    H=\frac{1}{2m}g^{\mu \nu}(p_\mu -qA_\mu)(p_\nu-qA_\nu),
\end{eqnarray}
where  $p_\mu$ stands for a generalized  covariant momentum
\begin{eqnarray}
p_\mu=mg_{\mu \nu}\dot{x}^\nu+qA_\mu.
\end{eqnarray}

In practice, it stems from one set of Hamilton canonical equations of the Hamiltonian system (9)
\begin{eqnarray}
    \dot{x}^\mu=\frac{\partial H}{\partial p_\mu}=\frac{1}{m}g^{\mu \nu}(p_\nu-qA_\nu).
\end{eqnarray}
$\dot{x}^\mu=(\dot{t},\dot{r},\dot{\theta},\dot{\varphi})$ represents a four-velocity,
which is a derivative of coordinate $x^\mu$ with respect to the proper time $\tau$.

Another set of Hamilton canonical equations of the Hamiltonian system (9) are
\begin{eqnarray}
    \dot{p}_\mu=-\frac{\partial H}{\partial x^\mu}=-\frac{1}{2m}g^{\alpha \nu}_{,\mu}(p_\alpha -qA_\alpha)(p_\nu-qA_\nu),
\end{eqnarray}
where $g^{\alpha \nu}_{,\mu}=\partial g^{\alpha \nu}/\partial x^\mu$.
They show the presence of two constants of motion
\begin{eqnarray}
p_t &=& -mf(r)\dot{t}=-E,\\
p_\varphi &=& mr^2\sin^2\theta \dot{\varphi }+q A_\varphi=L.
\end{eqnarray}

$E$ and $L$ correspond to the particle's specific energy $E$ and the $z$-direction angular momentum, respectively.

For simplicity,  dimensionless operations are given to the related variables and parameters
in terms of scale transformations: $t\to tM$, $\tau\to \tau
M$, $r\to rM$, $B\to B/M$, $E\to mE$, $p_t\to mp_t$, $p_r\to
mp_r$, $L\to mML$, $p_{\theta}\to mMp_{\theta}$, $q\to mq$, and
$H\to mH$. Thus, $m$ and $M$ in the above equations are two mass factors that can be
eliminated. In this way, the Hamiltonian (9) becomes a system of two degrees of freedom
\begin{eqnarray}
   H=\frac{1}{2} f(r) p_r^2 +\frac{p_\theta ^2}{2 r^2}-\frac{E^2}{2 f(r)}+\frac{\left(L-\frac{1}{2}
   \beta r^2 \sin ^2\theta \right)^2}{2 r^2 \sin ^2\theta},
\end{eqnarray}
where $\beta=qB$. For a time-like geodesic, the rest mass or the four-velocity $\dot{x}^\mu$
satisfying the relation $g_{\mu \nu}\dot{x}^\mu\dot{x}^\nu=-1$  corresponds to the conserved Hamiltonian
\begin{eqnarray}
    H=-\frac{1}{2}.
\end{eqnarray}
This is a third motion constant of the system (9).

If $\beta\neq 0$, there is no separation in the variables in the Hamilton--Jacobi equation of
the Hamiltonian system (15).
Therefore, the system (9) or (15) does not accept  a fourth motion constant and is nonintegrable. If $\beta=0$,
the fourth motion constant is admitted and the RGI Schwarzschild spacetime is dynamically integrable. These facts
show that the external asymptotically uniform magnetic field in Equation (6) does not
change the spacetime geometry (3)
but induces the nonintegrability of charged particle motions.

\section{Numerical investigations}

A time-transformed explicit symplectic integration algorithm is designed for the system (15). Then, it is used to
provide some insight into contributions of the dynamical parameters to chaotic dynamics.

\subsection{Setup of an explicit symplectic integrator}

As aforementioned, the Hamiltonian system (15) is nonintegrable. Numerical integration methods are very convenient to
solve such a nonintegrable system. Unlike traditional numerical methods such as
Runge--Kutta integrators, symplectic integrators exhibit good numerical performance
during a long-term integration of a Hamiltonian system.
They conserve the symplectic geometric structure and give no secular drift
to errors of motion constants during a long-term integration of a Hamiltonian system. In general,
explicit  symplectic integrators are superior to implicit ones at same orders in computational efficiency.
Most of the Hamiltonians from curved spacetimes have no separation of variables or are
not split into two explicitly integrable parts,
and thus do not seem to allow for the application of explicit  symplectic methods.  However,
more than two explicitly integrable splitting terms can exist in some curved spacetimes like
the Schwarzschild black hole spacetime, and explicit  symplectic methods are available without doubt~\cite{r32}.
Unfortunately, the Hamiltonian for the Kerr black hole metric are not directly split into
more than two explicitly integrable terms.
Nevertheless, this problem can be solved with the help of an appropriate time transformation to the Hamiltonian and explicit  symplectic methods can
still work well~\cite{r33,r34}. In what follows, we consider construction of  an explicit
symplectic integrator for the Hamiltonian system (15)
along the idea of time transformation.

Let the proper time $\tau $ be regarded as a new coordinate $q_0=\tau$ and its
corresponding momentum be $p_0$ with $p_0=-H=1/2\neq p_t$. We have
an extended phase-space Hamiltonian in the extended phase-space made of $(p_r,p_\theta,p_0; r,\theta,q_0)$
\begin{eqnarray}
    J =H+p_0.
\end{eqnarray}

Clearly, $J$ is always identical to zero, i.e., $J=0$, whenever the Hamiltonian  $H$ is conservative or not.
Choosing a time transformation
\begin{eqnarray}
    d\tau &=& g(r)dw, \\
    g(r) &=&  \frac{\gamma \Omega }{r^3}+\frac{ \Omega }{r^2}+1,
\end{eqnarray}
we have a new time transformation Hamiltonian
\begin{eqnarray}
   K &=& g(r)J \nonumber \\
    &=& g(r) \left(-\frac{E^2}{2 f(r)}+\frac{\left(L-\frac{1}{2} \beta    r^2 \sin ^2\theta\right)^2}{2   r^2 \sin ^2\theta}+p_0\right)\nonumber\\
   &&+\frac{g(r) p_\theta^2}{2 r^2}+\frac{1}{2} p_r ^2 \left(\frac{\gamma \Omega }{r^3}+\frac{\Omega }{r^2}-\frac{2}{r}+1\right).
   \end{eqnarray}

The time-transformed Hamiltonian $K$ has the following splitting pieces:
\begin{eqnarray}
    K=K_1+K_2+K_3+K_4+K_5+K_6,
\end{eqnarray}
where all sub-Hamiltonian systems are expressed as
\begin{eqnarray}
    K_1&=& g(r) \left(p_0-\frac{E^2}{2 f(r)}+\frac{\left(L-\frac{1}{2}\beta r^2 \sin^2\theta\right)^2}{2r^2\sin^2\theta}\right),\\
    K_2&=&\frac{p_{r}^2}{2},\\
    K_3&=&-\frac{p_{r}^2}{r},\\
    K_4&=&\frac{\Omega p_{r}^2}{2 r^2},\\
    K_5&=&\frac{\gamma \Omega  p_{r}^2}{2 r^3},\\
    K_6&=&\frac{g(r) p_\theta^2}{2 r^2}.
\end{eqnarray}

The six splitting parts have
analytical solutions that are explicit functions of the new time $w$.
Their solvers correspond to
$\mathcal{K}_1$, $\mathcal{K}_2$, $\mathcal{K}_3$, $\mathcal{K}_4$, $\mathcal{K}_5$,
and $\mathcal{K}_6$ in turn.
Setting $h$ as a time step, we have a second-order explicit
symplectic algorithm for the Hamiltonian (20)
\begin{eqnarray}
    S_2(h)&=& \mathcal{K}_6(\frac{h}{2})\times \mathcal{K}_5(\frac{h}{2})\times\mathcal{K}_4(\frac{h}{2})\times
    \mathcal{K}_3(\frac{h}{2})\nonumber\times\mathcal{K}_2(\frac{h}{2})\nonumber\\
    &&\times\mathcal{K}_1(h)\times\mathcal{K}_2(\frac{h}{2})\times\mathcal{K}_3(\frac{h}{2}) \times\mathcal{K}_4(\frac{h}{2})\times\mathcal{K}_5(\frac{h}{2})\nonumber\\
    &&\times\mathcal{K}_6(\frac{h}{2}).
\end{eqnarray}

Because the Hamiltonian (20) is equivalent to the Hamiltonian (15),
the  symplectic method (28) designed for the Hamiltonian (20)
is also suitable for the Hamiltonian (15). The solutions at the new time $w$
should be changed into those at the proper time $\tau$.

The integration time step is $h=1$. The parameters are $E=0.995$,
$L=4.1$,  $\beta=4\times 10^{-4}$, $\gamma =0.22$, and $\Omega=0.166$.  The initial
conditions are $p_{r}=0$ and $\theta=\pi/2$. The initial separation are $r=15$ for Orbit 1
and $r=92$ for Orbit 2. The initial value of $p_{\theta}~(>0)$ is given by
Equations (15) and (16). The method $S_2$ yields no secular growth in the Hamiltonian errors
$\Delta H=H+1/2$ of the two orbits in Figure 1a. This result is fit for a property of a symplectic integrator.
When the integration lasts $10^{7}$ steps, the proper time $\tau$ and the new time $w$ are almost the same,
as shown in Figure 1b.

In fact, the two tested orbits have different dynamical behaviors. Orbit 1 exhibits
a closed torus on the Poincar\'{e} section $\theta=\pi/2$ with $p_{\theta}>0$ in Figure 1c.
Such a torus is a characteristic of regular dynamics. Orbit 2 behaves as a thin area with
many random distribution points and therefore is chaotic. Why is the Hamiltonian error for the regular orbit 1
larger than that for  the chaotic orbit 2? It is because the regular orbit 1 has a smaller average period.
Although no period exists in the chaotic orbit 2, an average period $T$ is possible.
When $T$ increases, the Hamiltonian error with an order of $(h/T)^2$ decreases.

\subsection{Contributions of the parameters to chaotic dynamics}

As mentioned above, the technique of Poincar\'{e} sections is a good description of phase space structures
in the system (15) or (20). \textcolor{black}{Taking the parameters $E=0.995$, $L=4.5$, $\gamma=0.02$ and $\Omega=8.166$,
we plot Poincar\'{e} sections for three values of the magnetic parameter $\beta=$0.0004, 0.0008, 0.0012 in Figure 2a--c.
The plotted five orbits of charged particles around the RGI  Schwarzschild  black hole are regular for  $\beta=$0.0004 in Figure 2a.
Only one of the five orbits is chaotic for  $\beta=$0.0008 in Figure 2b. However, the five orbits are chaotic for  $\beta=$0.0012 in Figure 2c.
These facts show that the occurrence of chaos becomes easier as the magnetic parameter
increases from
a statistical viewpoint of the global phase space structure. The result on a larger value of the magnetic parameter
easily inducing chaos is also suitable for the Schwarzschild  case with  $\gamma=\Omega=0$ in Figure 2d--f.
Comparing the two cases of an RGI  Schwarzschild  black hole and a  Schwarzschild  black
hole, we find that
the five orbits are regular for $\beta=$0.0004
in panels a and d, but chaotic for  $\beta=$0.0012 in panels c and f. Chaos seems to be stronger in
panel f than in  panel c. In fact, an explicit difference between the two cases is that
for $\beta=$0.0008,  only one of the five orbits is chaotic in the RGI  Schwarzschild case in panel b, whereas
the five orbits are chaotic in the Schwarzschild case in panel e. Thus, for a given magnetic field parameter
under some circumstances, chaos seems to be much easily induced in the Schwarzschild black hole spacetime than in
the RGI Schwarzschild black hole spacetime. }

\textcolor{black}{Letting the parameters be $L=4.5$, $\gamma=0.089$, $\Omega=3.166$ and $\beta=8 \times 10^{-4}$,
we continue to consider the motion of charged particles around the RGI  Schwarzschild  black hole.
When the energies are given three values of $E=0.999$, $E=0.996$,  and  $E=0.991$ in
Figure 3a--c,
the decrease in the energy typically decreases the extent of chaos and even rules out the
occurrence of chaos.
However, the extent of chaos is typically strengthened when the particle's angular momentum $L$ decreases from
$L=6.5$ in Figure 3d to $L=5.5$ in Figure 3e and to $L=4$ in Figure 3f, where
the parameters are $\Omega=1.166$, $E=0.995$, $\gamma=0.089$ and $\beta=8 \times 10^{-4}$.
Similar results exist as the renormalization group parameter $\Omega$ decreases from
$\Omega=9.166$ in Figure 3g to $\Omega=2.166$ in Figure 3h and to $\Omega=0.166$ in Figure 3i, where
the parameters are $\beta=8 \times 10^{-4}$, $E=0.995$, $L=4.5$ and $\gamma=0.089$. }

The above-mentioned orbits
can be checked by  the maximal Lyapunov exponent, which was defined in Ref.~\cite{r35} as
\begin{eqnarray}
\lambda=\lim_{w\rightarrow\infty}\frac{1}{w}\ln\frac{d(w)}{d(0)},
\end{eqnarray}
where $d(w)$ is a proper distance of two nearby
orbits at the proper time $w$ and $d(0)$, one at the starting time.
$\lambda$ tending to zero shows regular dynamics for
$\beta=$0.0004 in Figure 4a, but two positive stabilizing values of $\lambda$
correspond to chaotic dynamics for $\beta=$0.0008, 0.0012. In particular, $\beta=$0.0012 with a larger Lyapunov exponent
leads to stronger chaos than $\beta=$0.0008. A quicker method to distinguish between order and chaos
is a fast  Lyapunov indicator (FLI) of Froeschl\'{e} and Lega~\cite{r36}. This indicator was redefined in terms of the two nearby-orbit method~\cite{r37}
as
\begin{eqnarray}
    FLI=\log _{10}\frac{d(w)}{d(0)}.
\end{eqnarray}

The FLIs that exponentially increase with time
$\log _{10}w$ describe the onset of chaos for $\beta=$0.0008, 0.0012 in Figure 4b,
while the FLI that algebraically grows with time
indicates the presence of order for $\beta=$0.0004. The methods of $\lambda$ and FLIs in Figure 4c--h
display that a larger energy or renormalization group parameter has larger Lyapunov exponent and FLI for a given orbit,
but a larger angular momentum corresponds to smaller Lyapunov exponent and FLI. Thus,
the effects of the parameters $E$, $L$, and $\Omega$
described by the methods of $\lambda$ and FLIs
are consistent with those given by the Poincar\'{e} map method in Figures 2~and~3.

\textcolor{black}{Taking the orbit with the initial radius
$r=40$ in Figure 2 as a tested orbit, we trace the FLI depending on the
magnetic parameter $\beta$. The FLI for a given value $\beta$ is obtained after
integration time $w=5\times10^6$. The FLIs $\geq$ 18.6 indicate
disorder dynamics, while the FLIs $<$ 18.6 show regular dynamics. In this way, the regular dynamics
and chaotic dynamics in the RGI Schwarzschild black hole spacetime are clearly shown through
the dependence of the FLI on the
magnetic parameter $\beta$ in Figure 5a.
The dynamical features in the Schwarzschild black hole spacetime are also drawn in Figure 5b.
The relations for the dependence of the FLI on the
magnetic parameter are nearly the same in the two spacetimes.
When the magnetic parameters are larger than some values, chaos can be allowed in the two spacetimes.
The increase in the magnetic parameter makes the chaoticity become easy and strong
under appropriate circumstances.
This result does not mean that a larger magnetic parameter must induce stronger chaos,
but is based on
most of the values of the magnetic parameter larger than some values from the statistical viewpoint of the global phase space structure.
However, there are some small differences between the two spacetimes. For $\beta\in [0.000713,0.001046]$,
regular dynamics occurs in the RGI Schwarzschild case, but chaotic dynamics does in the Schwarzschild case.
For $\beta\in [0.001847,0.001928]$, the RGI Schwarzschild case allows for chaos, while the Schwarzschild case does not. }

Using the FLIs corresponding to two parameter spaces, we detect chaos from order in the
two parameter spaces.
For example, the initial radius is $r=40$ and the FLIs are obtained from the two-parameter
space $(\Omega,\beta)$,
where $0.165\leq \Omega\leq 9.804$ and $0.0002\leq\beta\leq0.0018$. It can be seen clearly from  Figure 6a that
an increase in $\beta$ enhances the strength of chaos, while an increase in $\Omega$
weakens the strength of chaos.
Such an effect of varying the renormalization group parameter $\Omega$ on
the dynamical transition of particles from order to chaos also exists in the two-parameter
spaces $(\Omega,E)$ and $(\Omega,L)$ in Figure 6b,c.
In addition, chaos occurs easily with the energy $E$ increasing, but does not  with the angular momentum $L$ increasing.
These results are also similar to those in Figures 2 and 3.

Now, let us consider the impact of the distance scale $\gamma\in[0.02,0.22]$ on the dynamical transition.
When the starting separation $r=40$ and the
parameters $E=0.995$, $L=5.157$, $\beta=8\times 10^{-4}$ and \textcolor{black}{$\Omega=0.166$} are given, three values
$\gamma=0.08$, 0.12, and 0.19 allow for the existence of chaos, as shown via the methods
of Poincar\'{e} sections, Lyapunov exponents,
and FLIs in Figure 7a-c. The dynamical transition is insensitive dependence on a variation of  $\gamma\in[0.02,0.22]$ in the parameter spaces
$(\gamma,\beta)$, $(\gamma, E)$, and $(\gamma,L)$ of Figure 6d-f.
The dynamical transition insensitive dependence is based on the parameter $\gamma$
constrained in Ref.~\cite{r11}. However, the  ISCO  radius increasing with
the increase in the parameter $\gamma$ in Ref.~\cite{r12} is due to the parameter
$\gamma$ dissatisfying this constraint.

The main results can be concluded from Figures 2-7. Under
some circumstances, chaos becomes easier as the energy $E$ and the magnetic parameter
$\beta$ increase
but does not when  the angular momentum $L$ and the renormalization group parameter $\Omega$ increase.
A small variation in the distance scale  $\gamma$ seems to exert a negligible influence on
the dynamical transition.
In fact, these results are obtained from the statistical viewpoint of the global phase space structure.
In order to explain them, we expand the third and fourth terms of Equation (15) outside
the horizons as
\begin{eqnarray}
   H_1 &=& -\frac{E^2}{2 f(r)}+\frac{\left(L-\frac{1}{2}
   \beta r^2 \sin ^2\theta \right)^2}{2 r^2 \sin^2\theta}, \nonumber \\
   &\approx & -\frac{E^2+\beta L}{2} -\frac{E^2}{r}+\frac{1}{8}\beta^2r^2\sin^2\theta- \frac{E^2}{r^2}+\frac{L^2}{2r^2\sin^2\theta}  \nonumber \\
   && +\frac{E^2\Omega}{r^3}-\frac{E^2\Omega}{r^4}(\gamma-2\Omega)+\cdots.
   \end{eqnarray}

The second term presents that the black hole gives an attractive force contribution to the particle motion.
The third term acts as a magnetic field  attractive force and causes chaos to easily occur.
The sixth term and the fifth term yield  a repulsive force effect, which
can somewhat suppress chaos. If $\gamma\ll 2\Omega$ and  $\gamma\in[0.02,0.22]$, the contribution of $\gamma$ to  the seventh term seems is negligible;
if $\gamma>2\Omega$ and  $\gamma\in[0.02,0.22]$, the seventh term is still
denominated by the sixth term.
Why would a smaller value of the renormalization group parameter much easily induce chaos than a larger value? The question
can be answered with the help of
the universal relations between the chaotic test particle motions around the black holes and the surface gravities of such black hole horizons~\cite{r38,r39}.
Considering the motion of a particle near a horizon of a spherically symmetric black hole, the authors of [38] obtained a
universal Lyapunov exponent provided by the surface gravity of the black hole. The authors of [39] estimated
the radial distance of a massless and chargeless particle very near to the event horizon having exponential growing nature.
For simplicity, we take $\gamma=0$ in the present black hole geometry (2) and (3). The surface gravity of the black hole
is expressed in [38] as
\begin{eqnarray}
 \kappa=\sqrt{\frac{-g_{tt}}{g_{rr}}}\frac{d}{dr}\left( \ln \sqrt{-g_{tt}} \right)=\frac{1}{2}f'(r).
   \end{eqnarray}
Given $\Omega=0$, the spacetime (3) is the Schwarzschild black hole with the  horizon $r_{H}=2$.
In the case, the surface gravity near the horizon is
\begin{eqnarray}
 \kappa=\frac{1}{r^2_H}=\frac{1}{4}.
   \end{eqnarray}
When $\Omega=1$, the spacetime (3) also has a  horizon $r_{H}=1$, and the surface gravity near the horizon is
\begin{eqnarray}
 \kappa=\frac{1}{2r^2_H}\left(r_H-1\right)=0.
   \end{eqnarray}
If $0<\Omega<1$, the black hole geometry (3) has two horizons $r_H=r^{\pm}_H=1\pm\sqrt{1-\Omega}$,
and the surface gravity near the horizon $r=r^+_H$ is
\begin{eqnarray}
 \kappa=\frac{1}{2r^{+2}_H}\left(r^+_H-\Omega\right)>0.
   \end{eqnarray}
That is, this inequality is always satisfied for the case of $0<\Omega<1$.
However, $\kappa<0$ near the horizon $r=r^-_H$.
In fact, the surface gravity of the black hole  near the horizon was defined as a Lyapunov exponent $\lambda_H=\kappa$ [38].
Thus, the Lyapunov exponent $\lambda_H$ near the horizon $r=r_{H}$ or $r=r^+_H$  is positive for the case of $0\leq\Omega<1$.
The horizon would influence chaotic behavior in the motion of charged particles in the Hamiltonian system (15).
Although this analysis is based on the case of  $\gamma=0$, it is also suitable for the case of  $\gamma\neq0$
because the small distance scale $\gamma$ almost has a negligible influence on the dynamics.

\section{Summary}
\label{sec1}

An RGI Schwarzschild black hole metric is a nonrotating black hole solution in the
quantum gravity as a kind of  modified gravity theory.
In this metric, the Newton's gravitational constant is modified as
a position-dependent quality. There are also two spatio-temporal quantum correction
parameters, including the identification of cutoff of the distance scale and a nonperturbative renormalization group theory.
No black hole singularity exists due to the quantum gravity effects from the loop quantum gravity and the RGI
approach. The RGI Schwarzschild black hole spacetime is similar to the Schwarzschild black hole spacetime that is integrable.
In other words, the motion equations of test particles around the RGI Schwarzschild black hole have formally analytical solutions,
and the particles' motion is regular.

When an external asymptotically  uniform magnetic field surrounds the RGI Schwarzschild black hole, it is so weak that it does not change
the spacetime geometry. However, it strongly affects the motion of electrically charged test particles near the RGI Schwarzschild black hole
when the ration of charge to mass is large. Even the motion of charged particles is nonintegrable and may allow for the occurrence of chaos.
A time-transformed explicit second-order symplectic integrator exhibits good long-term performance
and effectively simulates the dynamical evolution of the nonintegrable system.

Using the methods of Poincar\'{e} sections, Lyapunov exponents  and the FLIs, we investigate a varying parameter how to affect the dynamical transition
from regular dynamics to chaotic dynamics.
From a statistical viewpoint, chaos occurs easily under some circumstances
as the specific energy and the magnetic parameter increase,
but does not when  the angular specific momentum and the renormalization group parameter $\Omega$ increase.
A small variation in the distance scale has a negligible effect on the dynamical transition. An interpretation is given to the
dependence of the dynamical transition on one varying
parameter. In particular, a smaller value of the renormalization group parameter would much easily induce chaos than a larger value
because of  a positive Lyapunov
exponent that is universally given by the surface gravity of the black hole when $\Omega\geq 0$ is small and the external magnetic field vanishes.
In this case, the horizon would influence chaotic behavior in the motion of
charged particles around the black hole surrounded by the external  magnetic field.


\textbf{Author Contributions}: J.L. made contributions to the software, and writing-original draft.
X.W. contributed to the supervision, conceptualization, writing-review and editing, and
funding acquisition. All authors have read and agreed to the published version of the manuscript.

\textbf{Funding}: This research was supported by the National Natural Science Foundation of China (Grant
No. 11973020).

\textbf{Data Availability Statement}: All of the data are shown as the figures and formula. No
other associated data.

\textbf{Conflicts of Interest}: The authors declare no conflict of interest.

\begin{figure*}[htpb]
        \centering{
        \includegraphics[width=13pc]{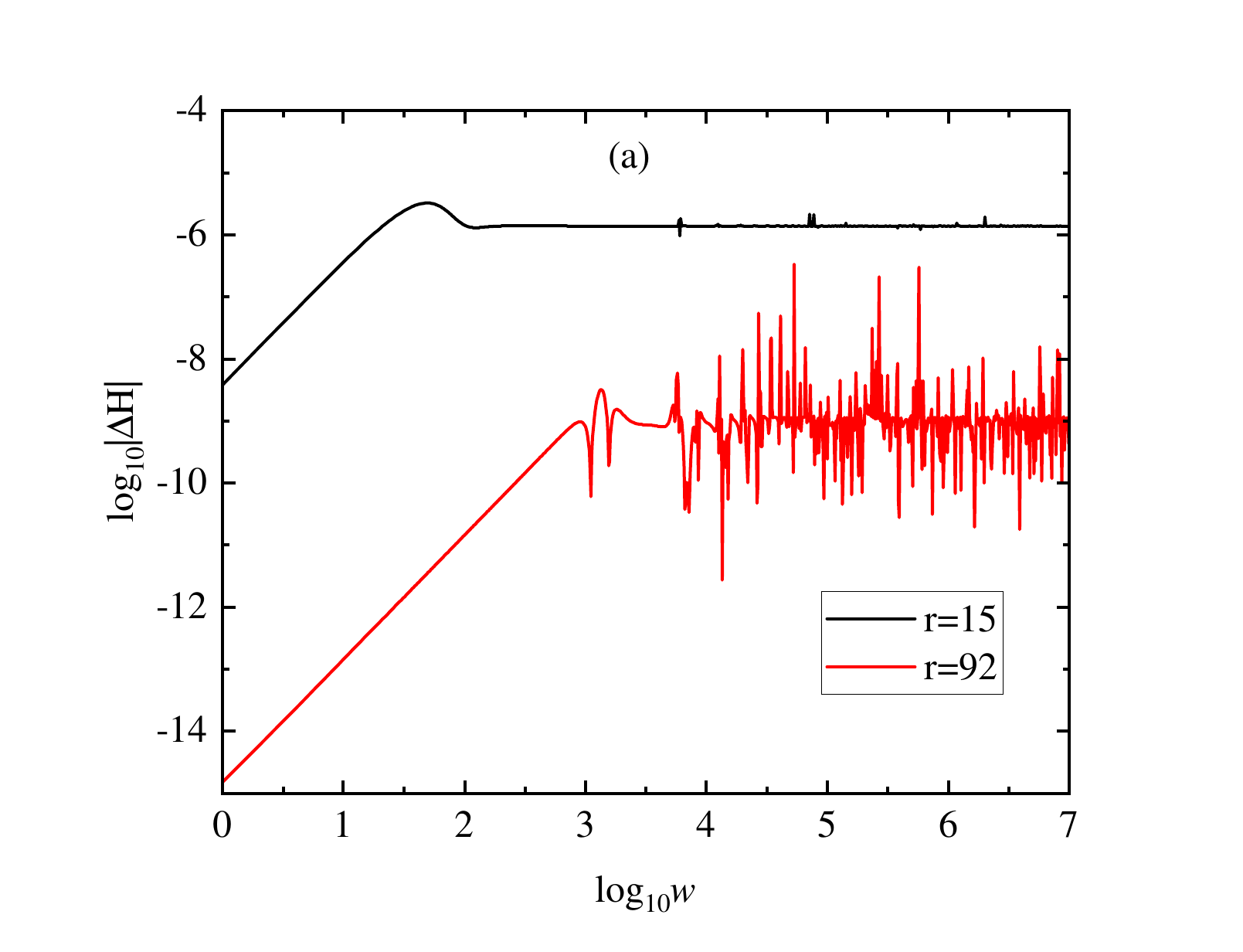}
        \includegraphics[width=13pc]{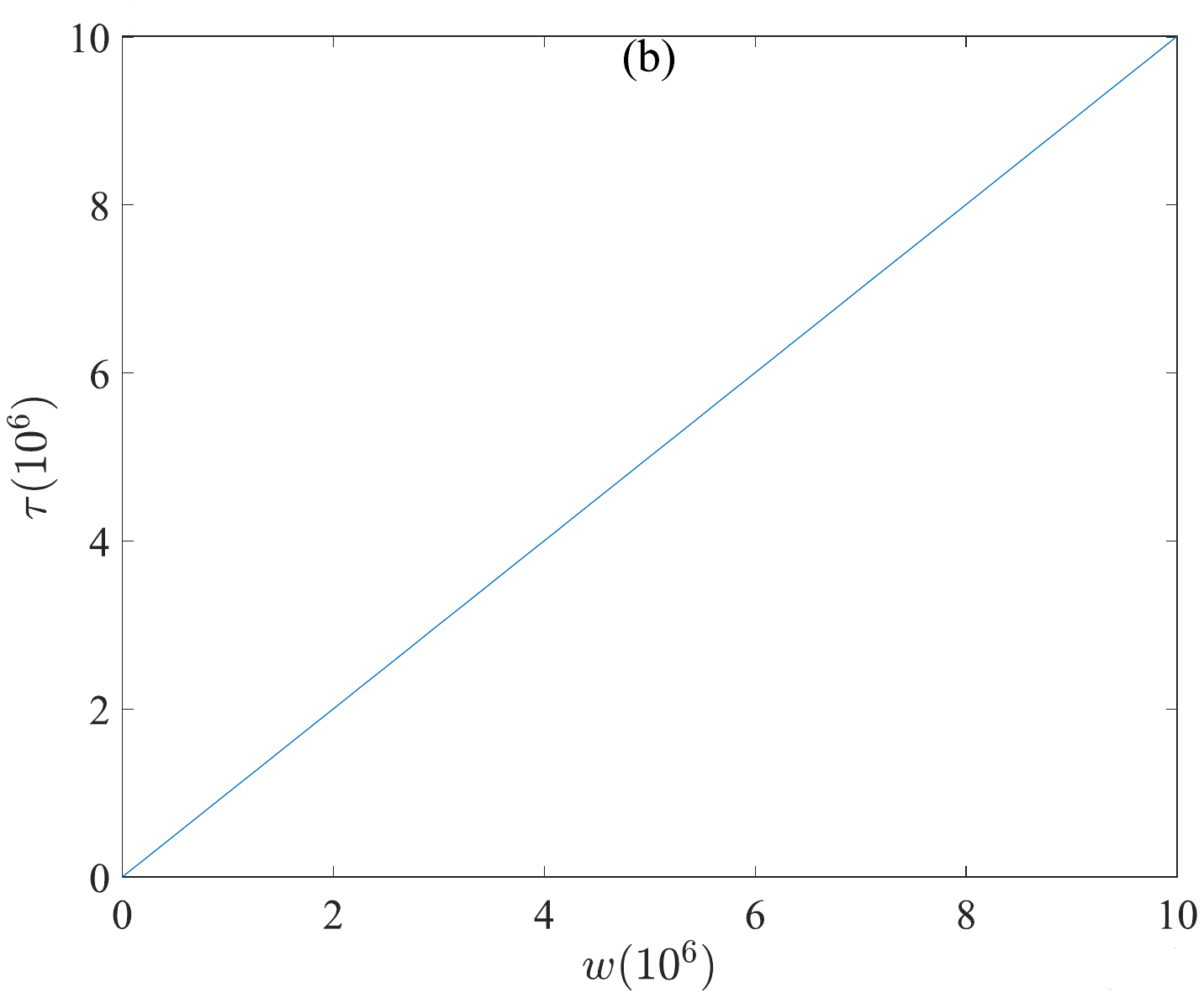}
        \includegraphics[width=13pc]{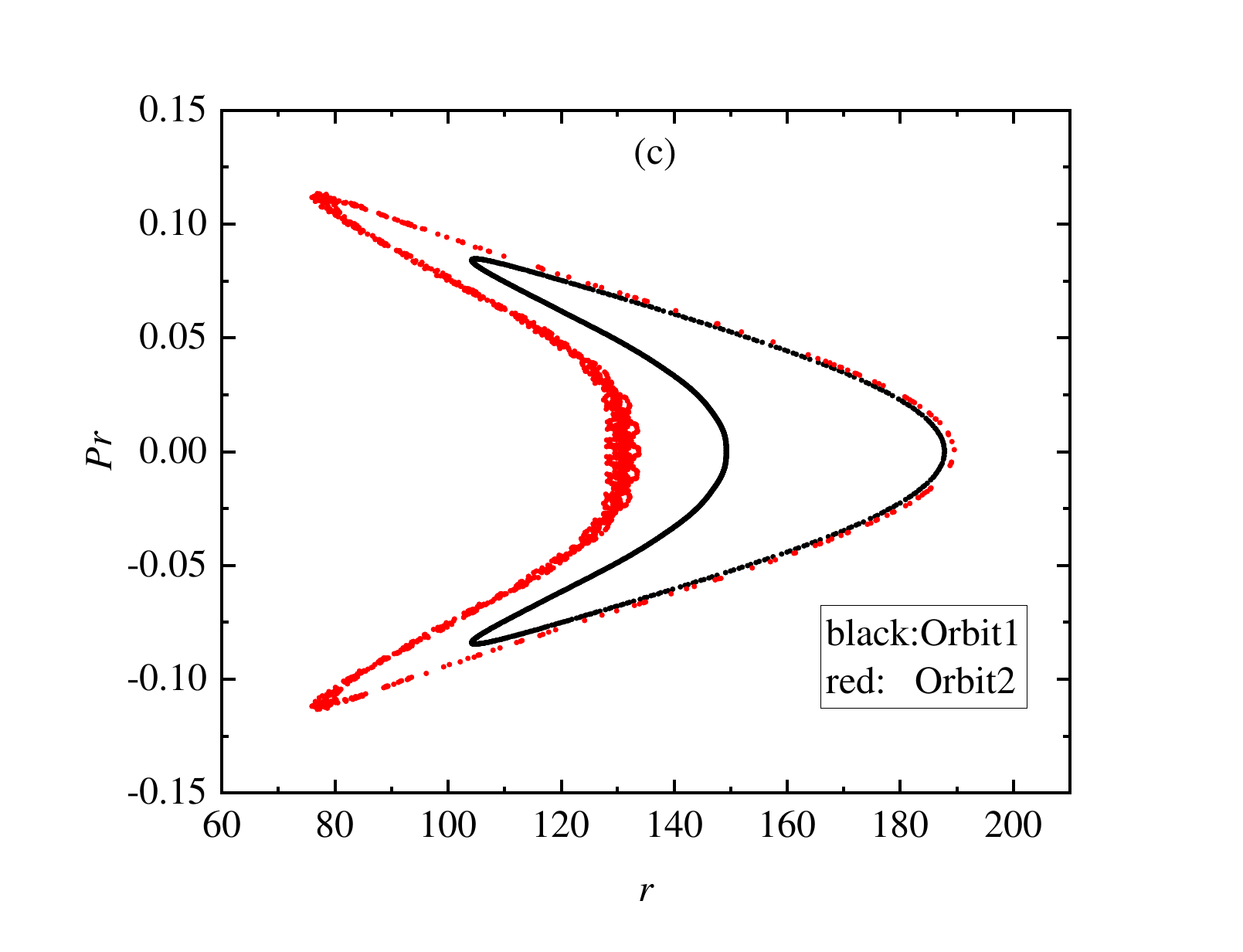}
\caption{  (a) The Hamiltonian errors $\Delta H=H+1/2$ for two orbits solved by the explicit symplectic integrator with step size $h=1$.
The parameters are $E =0.995$, $L = 4.1$, $\beta =4\times 10^{-4}$, $\gamma=0.22 $, and $\Omega = 0.166$.
The initial conditions are $p_r=0$ and $\theta=\pi/2$, and the initial separations are $r=15$ for Orbit 1 colored Black  and $r=92$ for Orbit 2 colored Red.
The errors show no secular drift.
(b) The relation between proper time $\tau$ and the new time $w$. The two times are nearly identical.
(c) Poincar\'{e} sections of the two orbits at the plane $\theta=\pi/2$ with $p_{\theta}>0$. Orbit 1 is regular, while Orbit 2 is chaotic.
        }
    }
\end{figure*}
\begin{figure*}[htpb]
    \centering{
        \includegraphics[width=13pc]{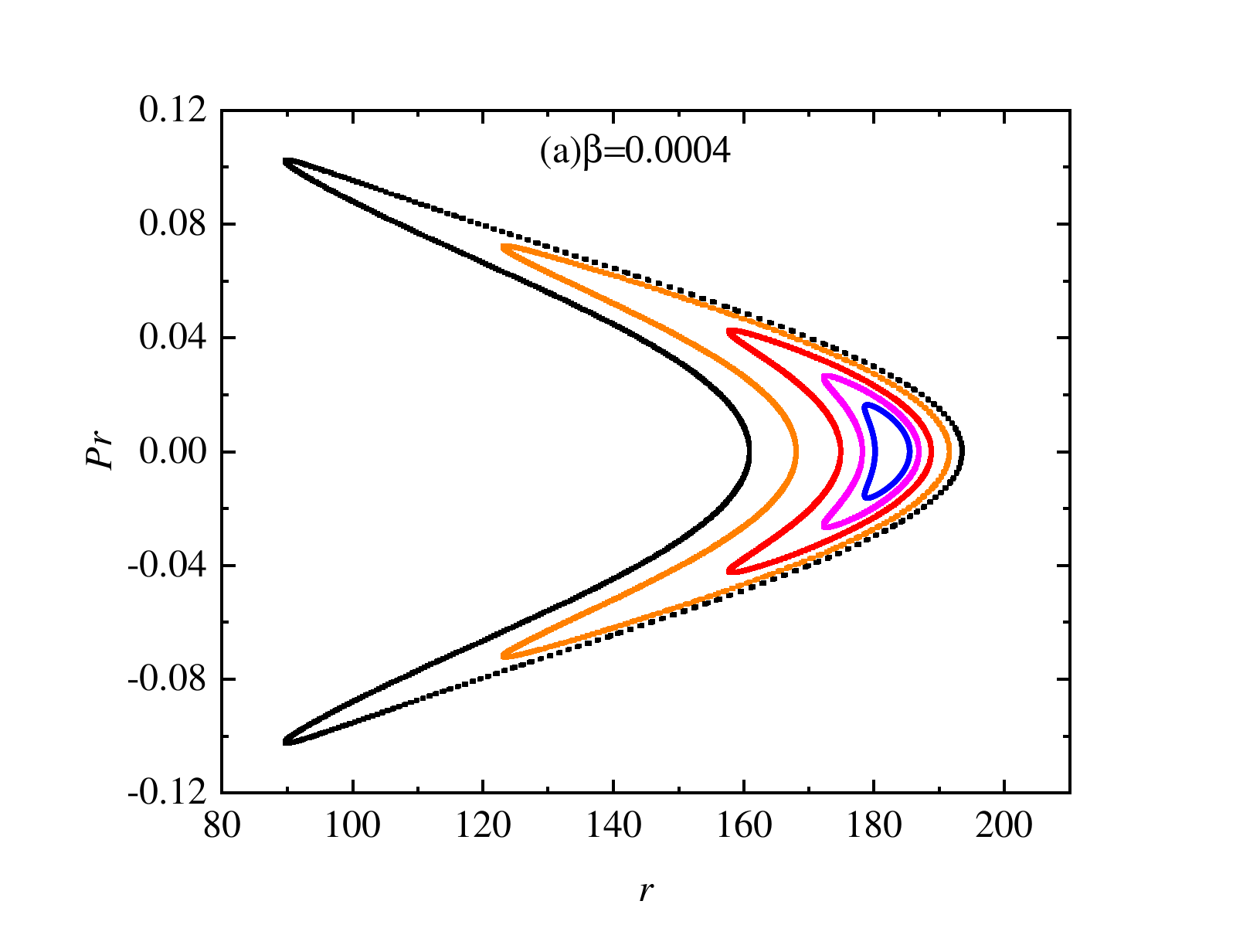}
        \includegraphics[width=13pc]{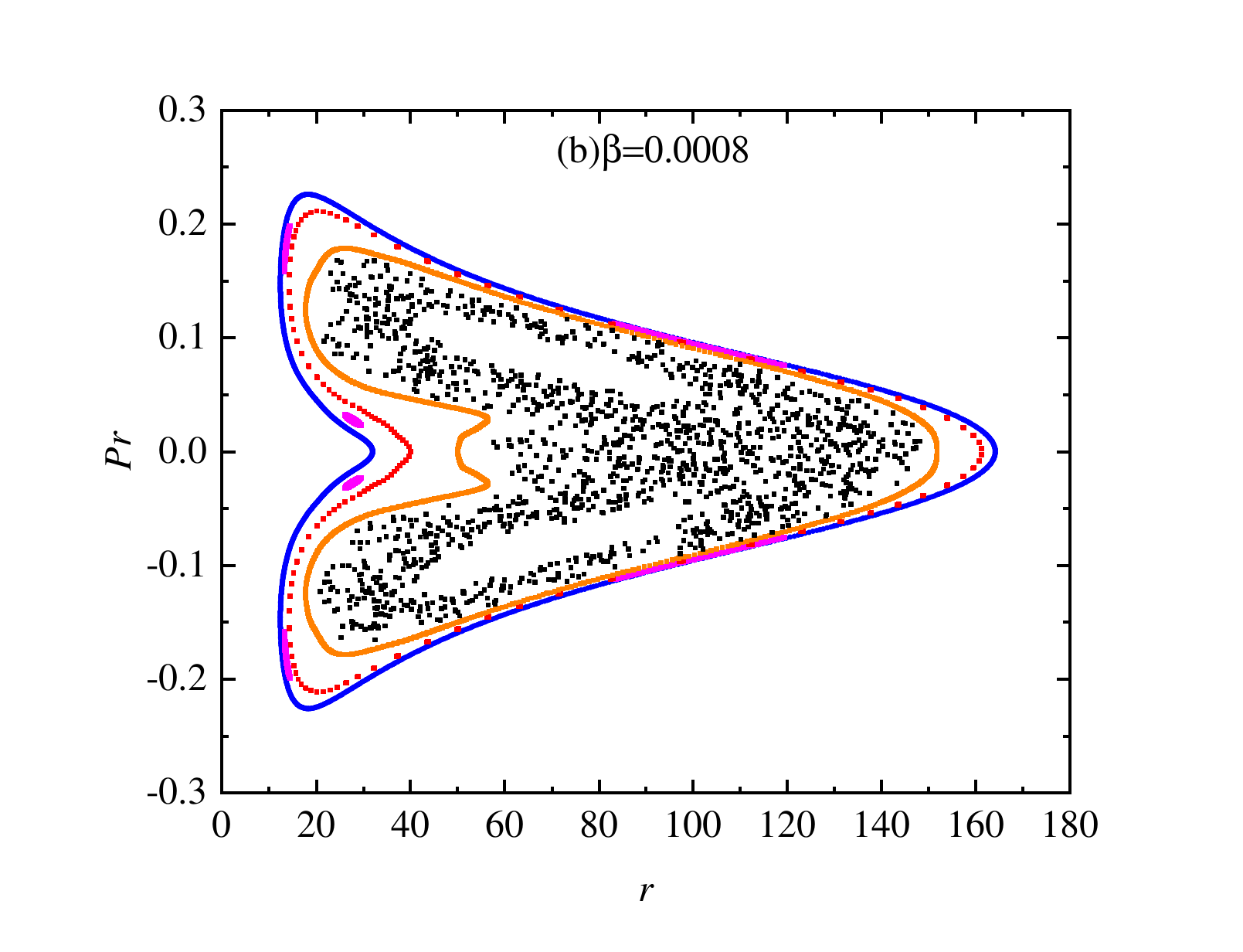}
        \includegraphics[width=13pc]{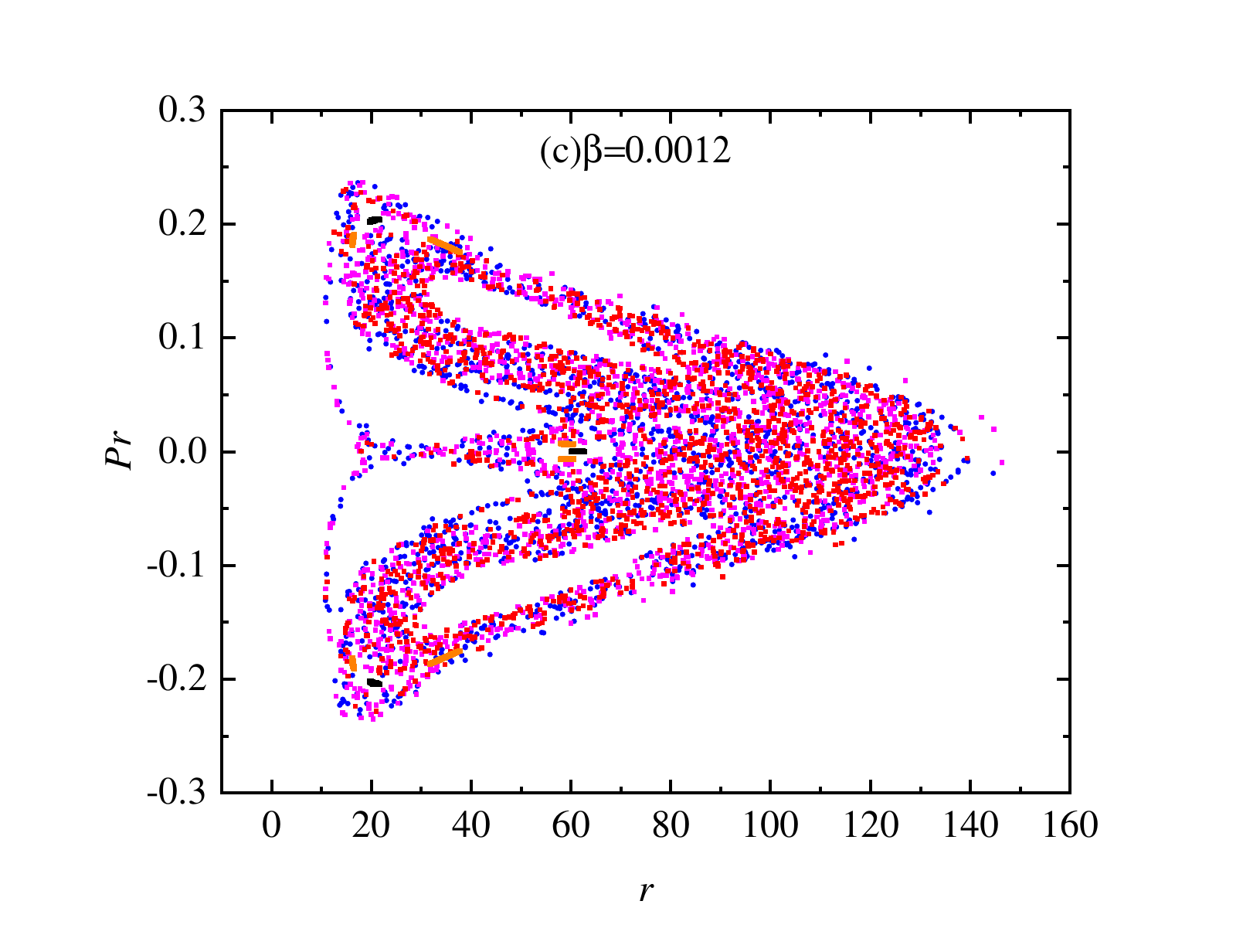}
        \includegraphics[width=13pc]{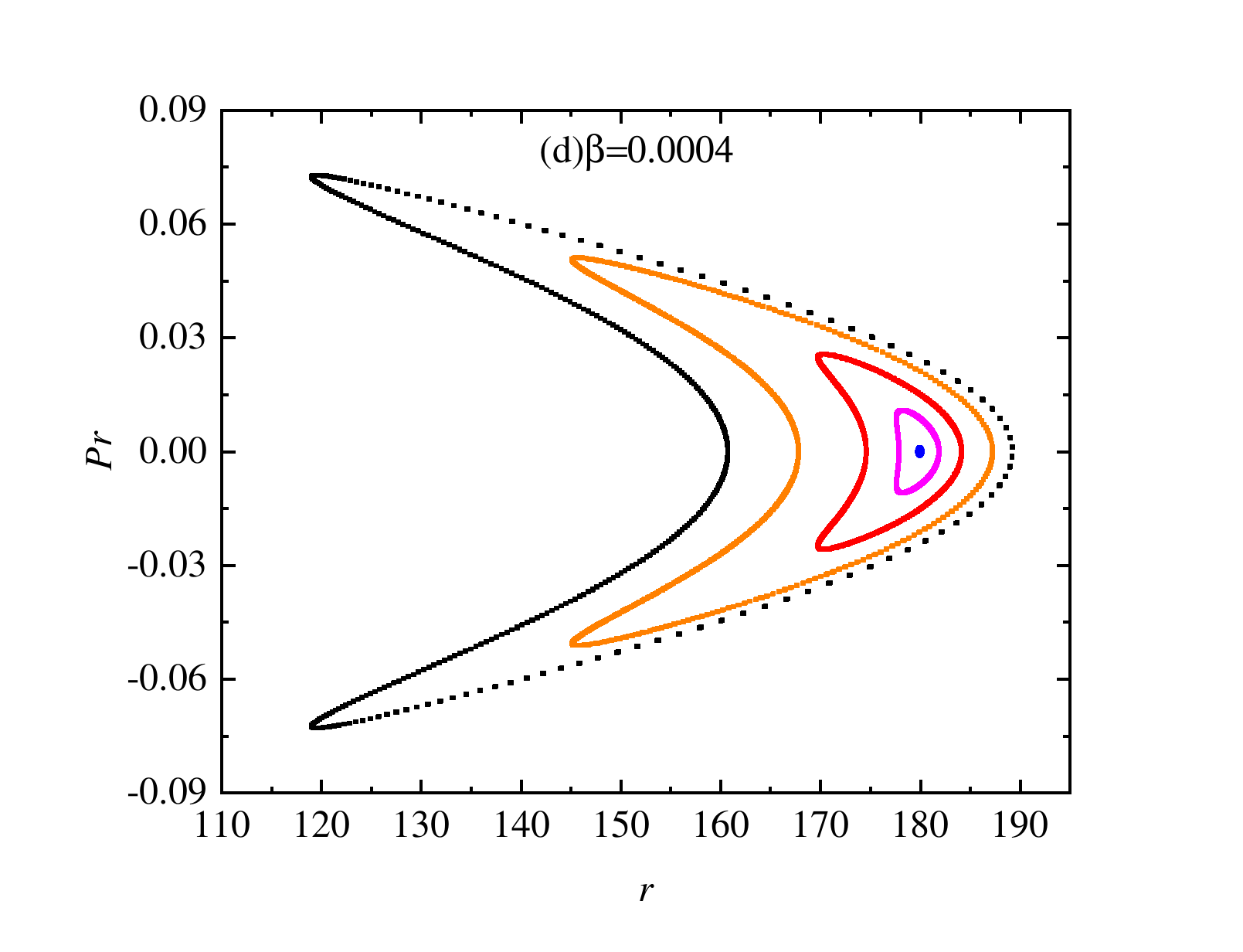}
        \includegraphics[width=13pc]{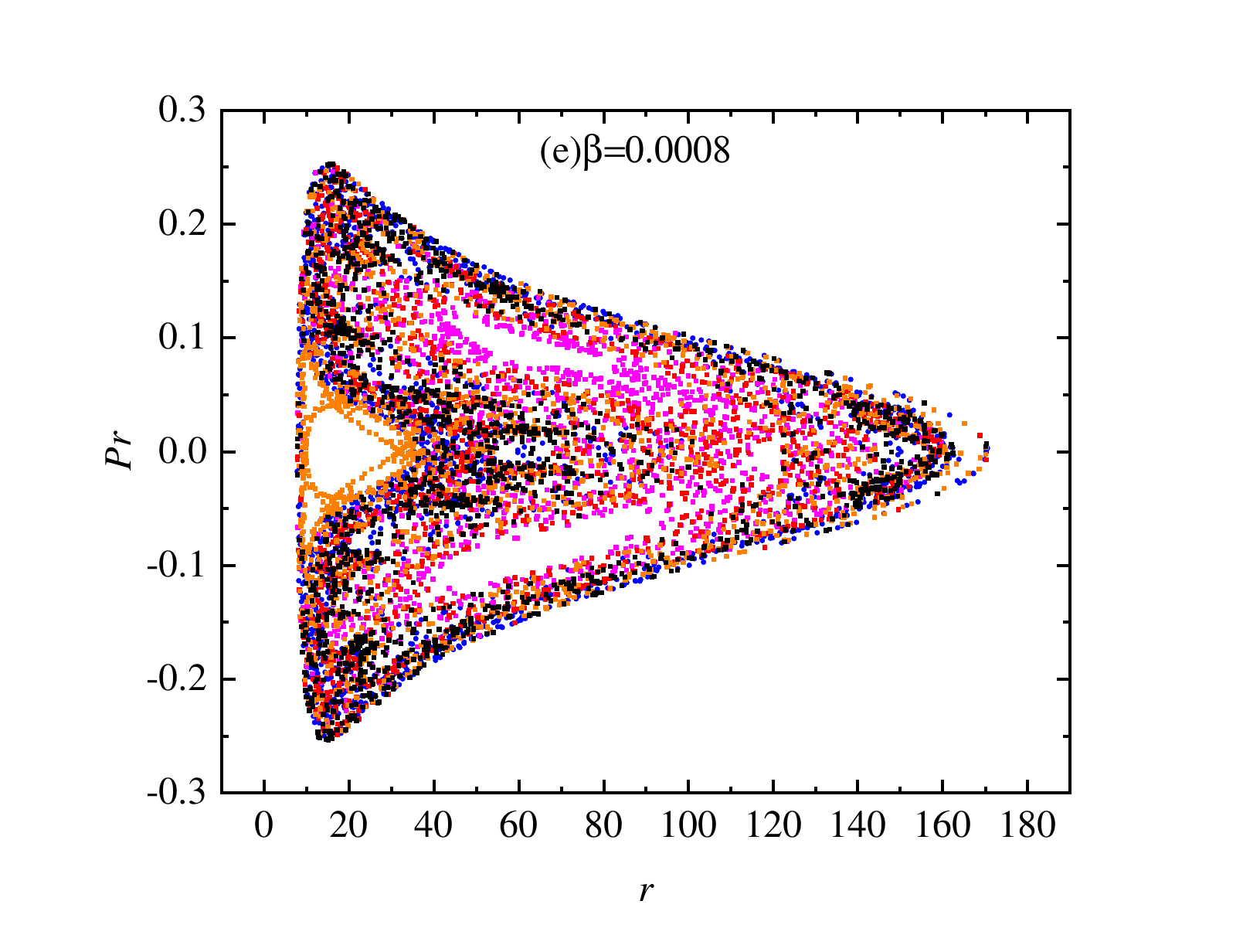}
        \includegraphics[width=13pc]{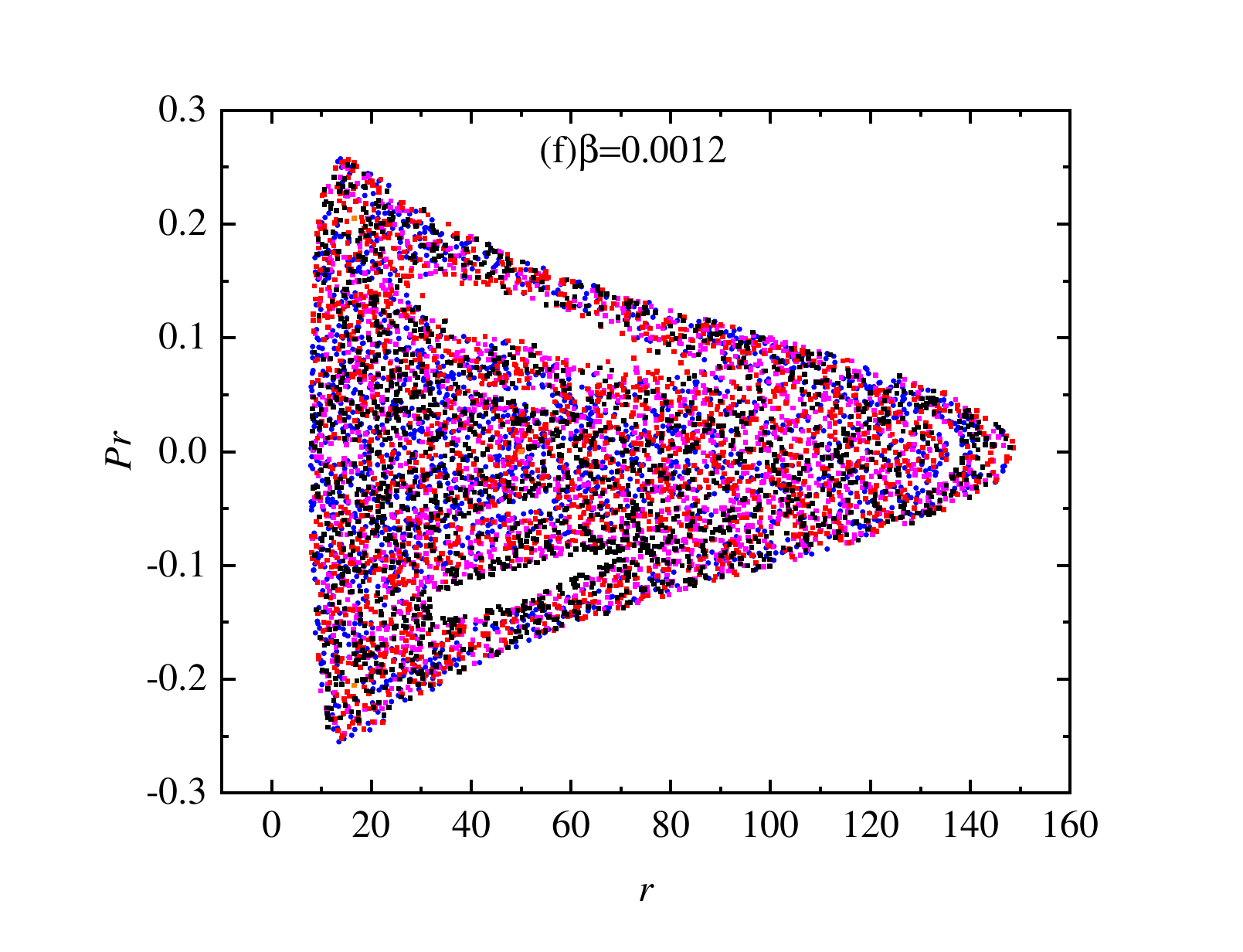}

       \caption{Poincar\'{e} sections. The parameters are $E=0.995$ and $L=4.5$. In the RGI Schwarzschild spacetime with $\gamma=0.02$ and $\Omega=8.166$, the magnetic field parameters are (a) $\beta=0.0004$, (b) $\beta=0.0008$, and (c) $\beta=0.0012$.
    In the Schwarzschild spacetime with $\gamma=\Omega=0$, the magnetic field parameters are (d) $\beta=0.0004$, (e) $\beta=0.0008$, and (f) $\beta=0.0012$.
   Clearly, chaos occurs easily in the two spacetimes as the magnetic field parameter increases.  For $\beta=0.0004$, the dynamical
   structure without the onset of chaos in the RGI Schwarzschild spacetime in panel (\textbf{a}) is similar to that in the Schwarzschild spacetime in panel (d).
  For $\beta=0.0008$  in panels (b) and (e), more orbits can be chaotic in the Schwarzschild spacetime.  For $\beta=0.0012$,
  chaos becomes stronger in the Schwarzschild spacetime in panel (f) than in the
  RGI Schwarzschild spacetime in panel (c).
        }

    }
\end{figure*}
\begin{figure*}[htpb]
    \centering{
        \includegraphics[width=13pc]{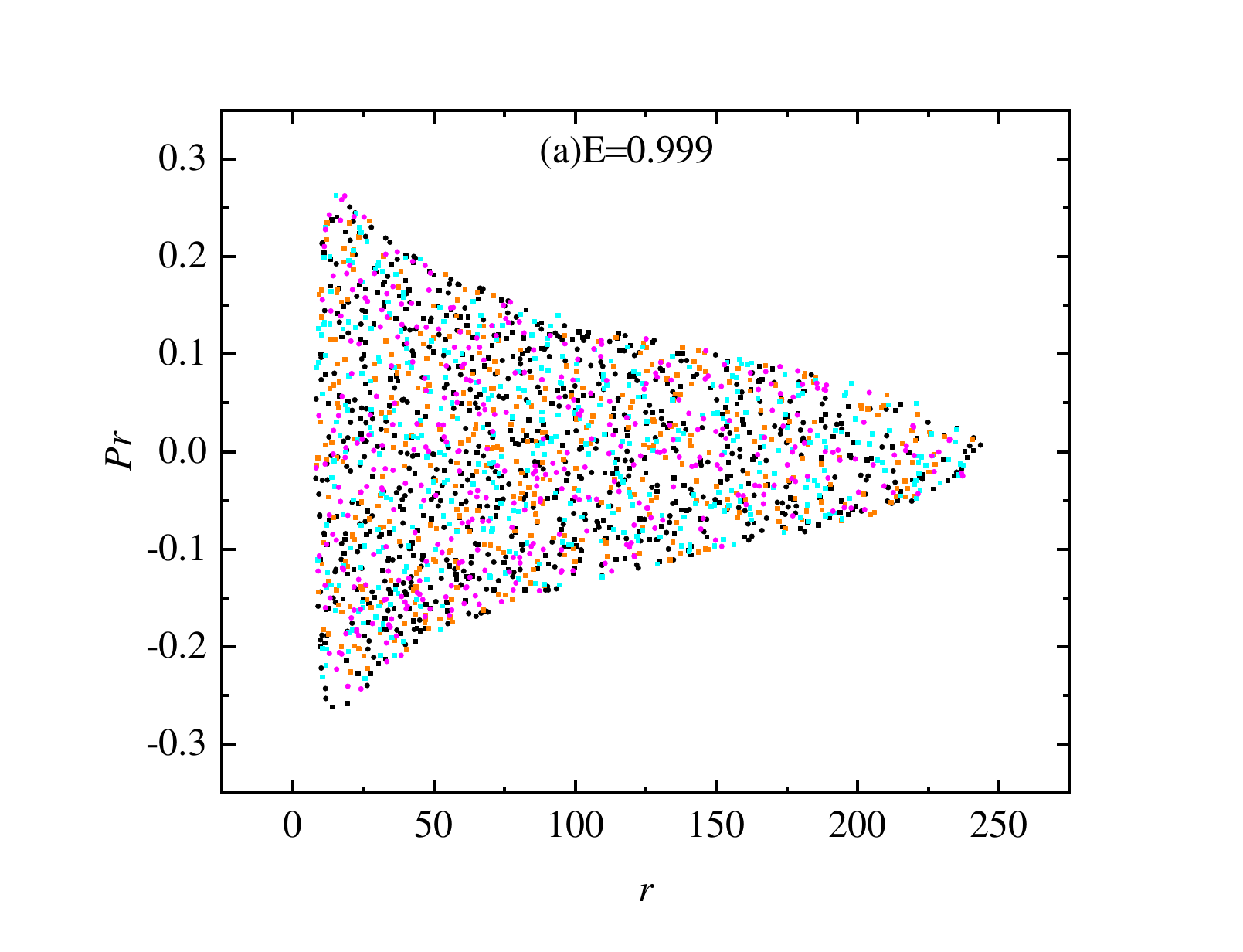}
        \includegraphics[width=13pc]{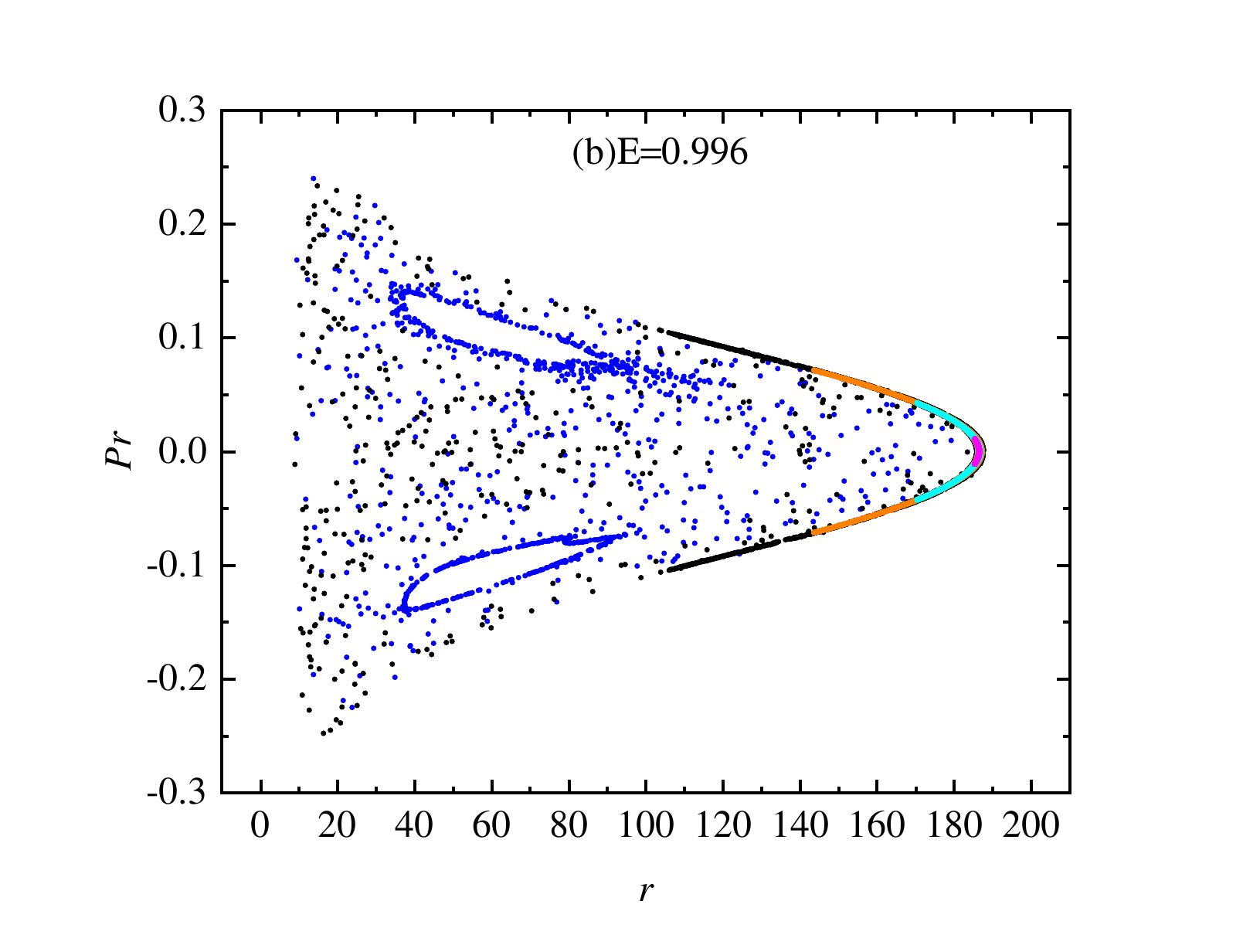}
        \includegraphics[width=13pc]{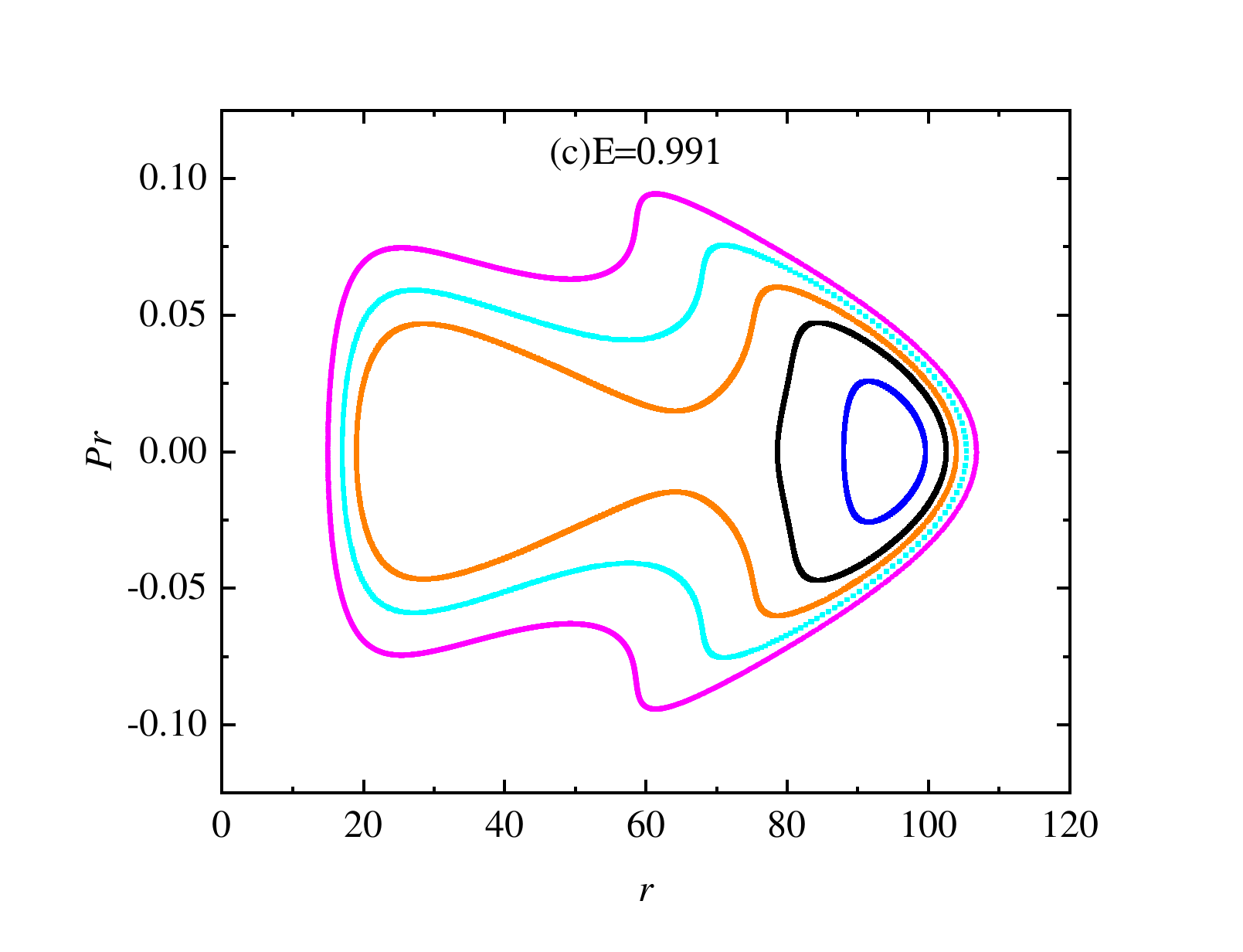}
        \includegraphics[width=13pc]{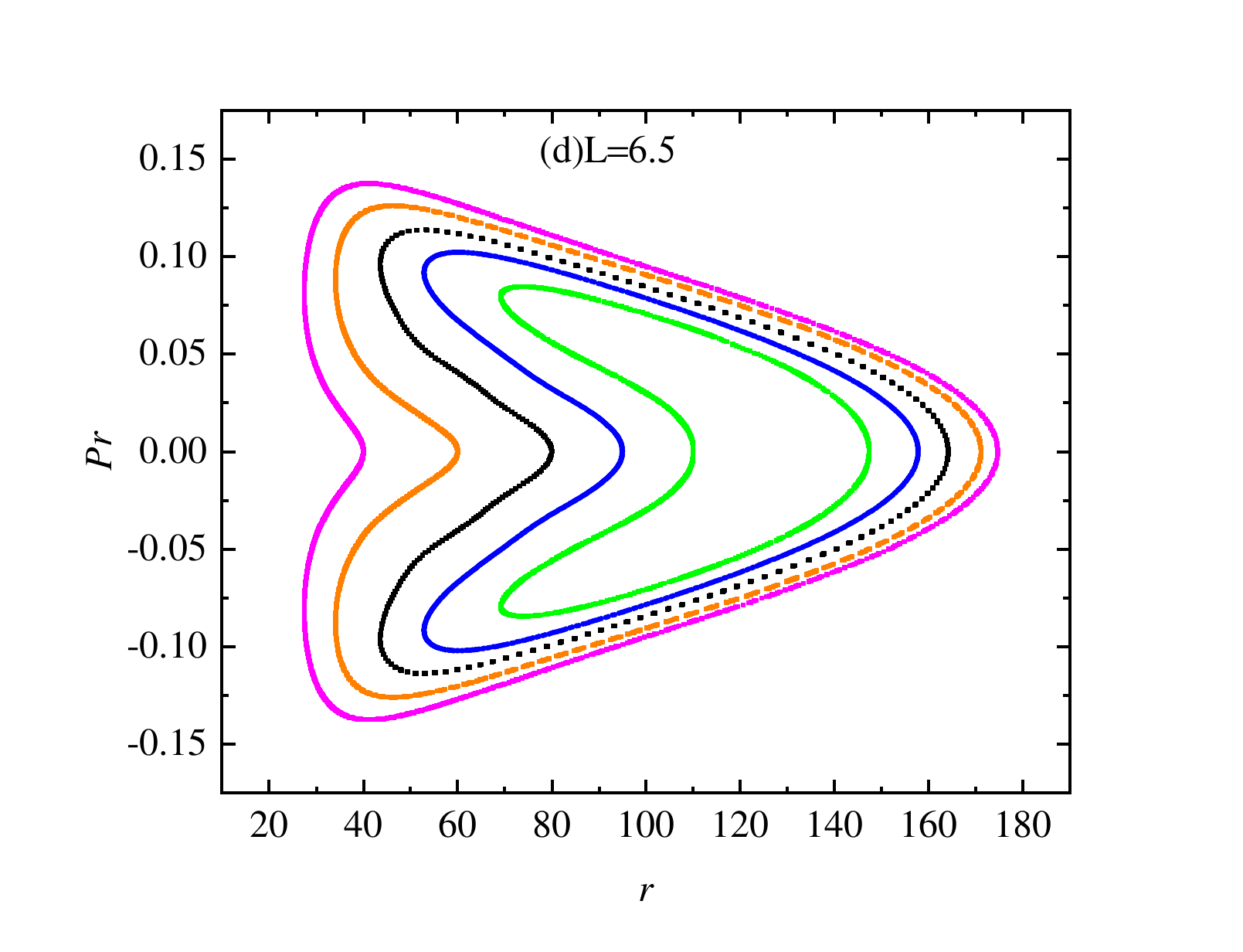}
        \includegraphics[width=13pc]{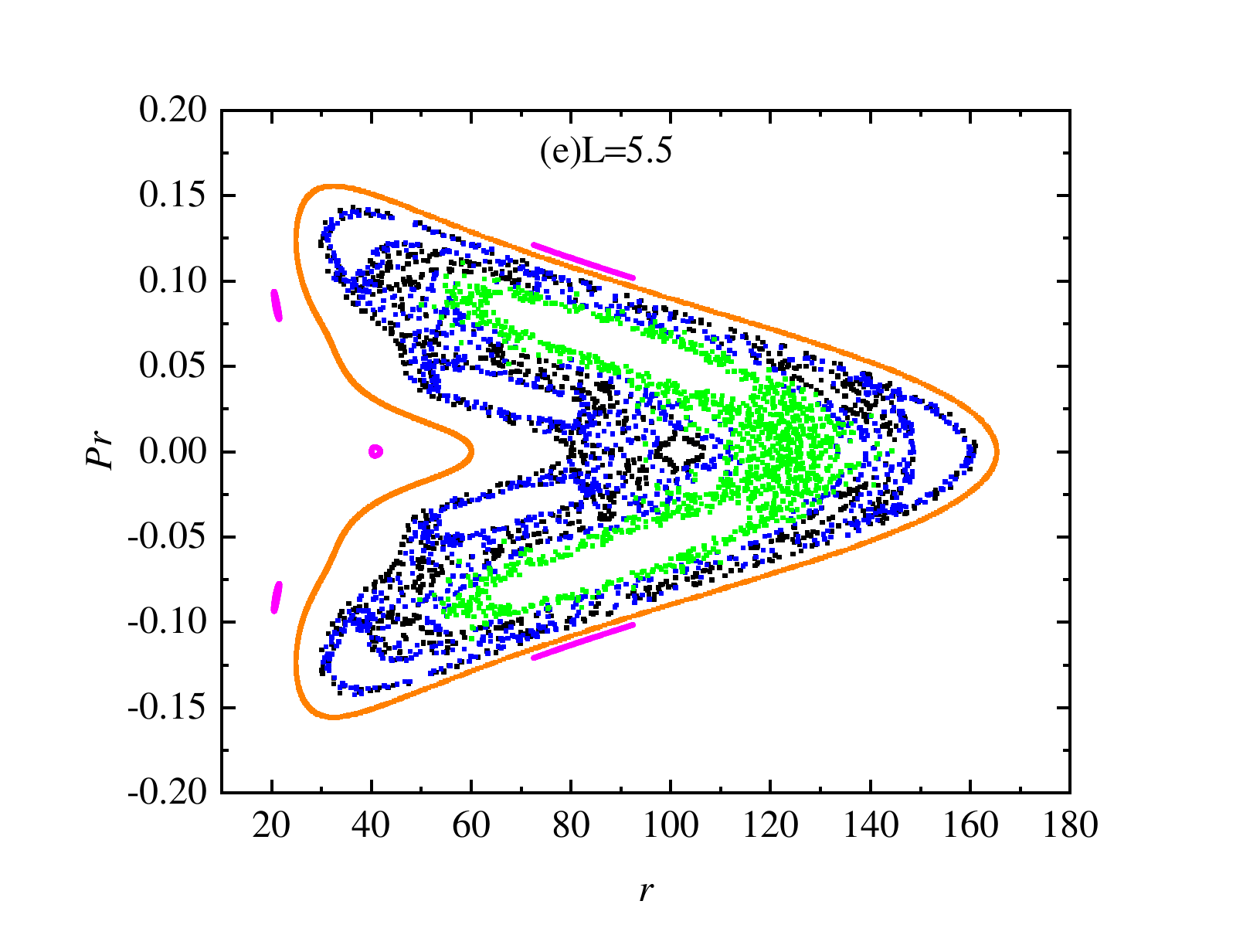}
        \includegraphics[width=13pc]{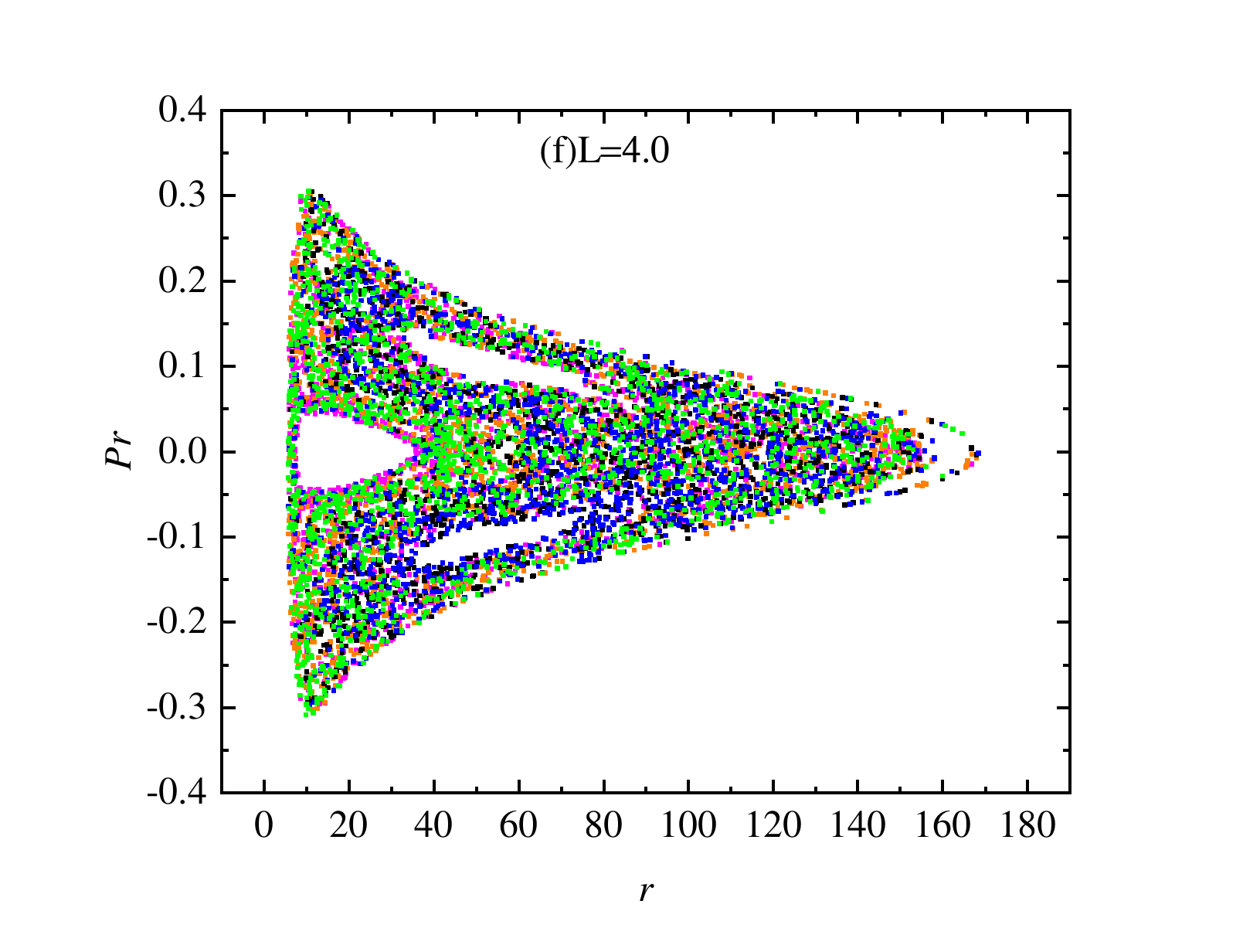}
        \includegraphics[width=13pc]{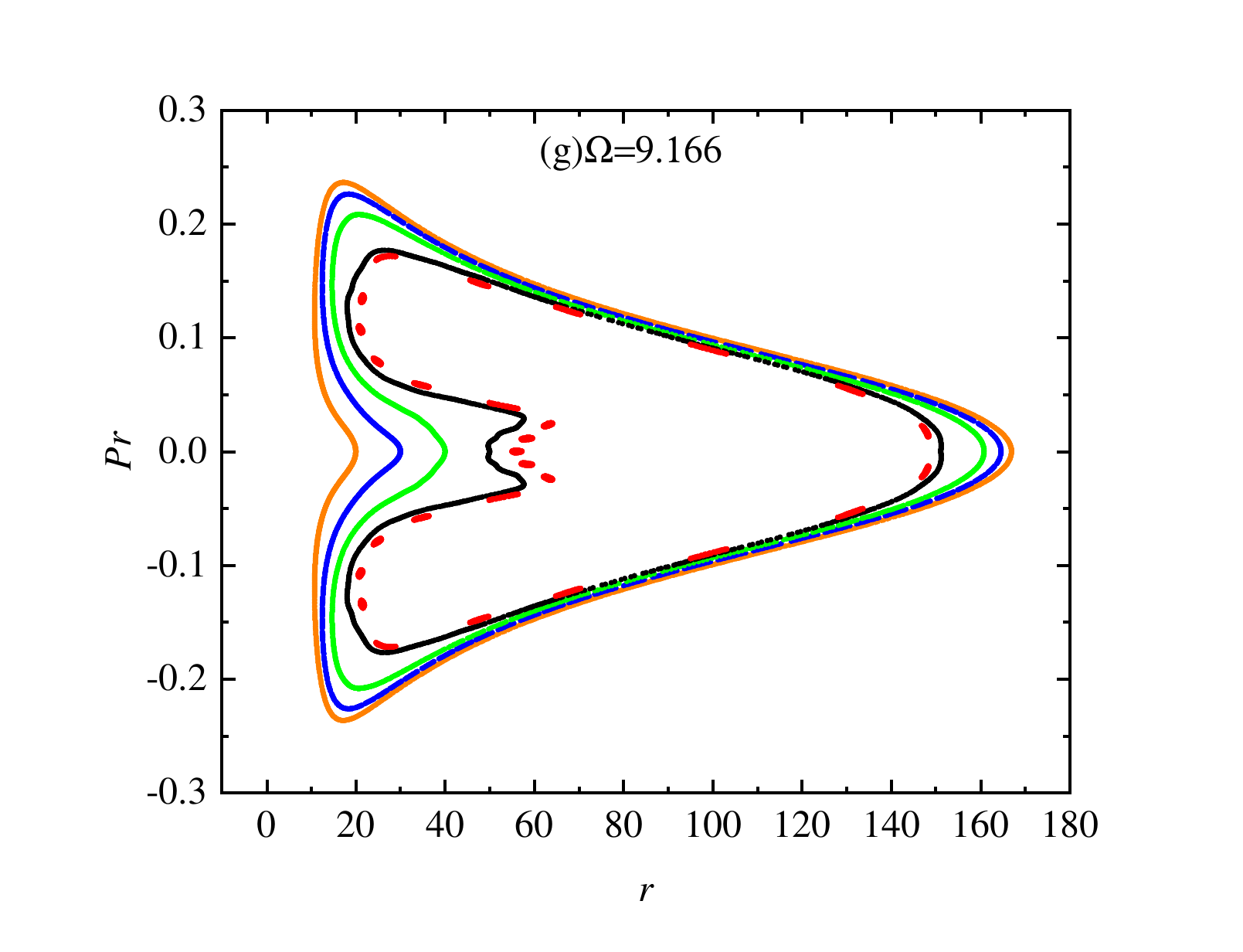}
        \includegraphics[width=13pc]{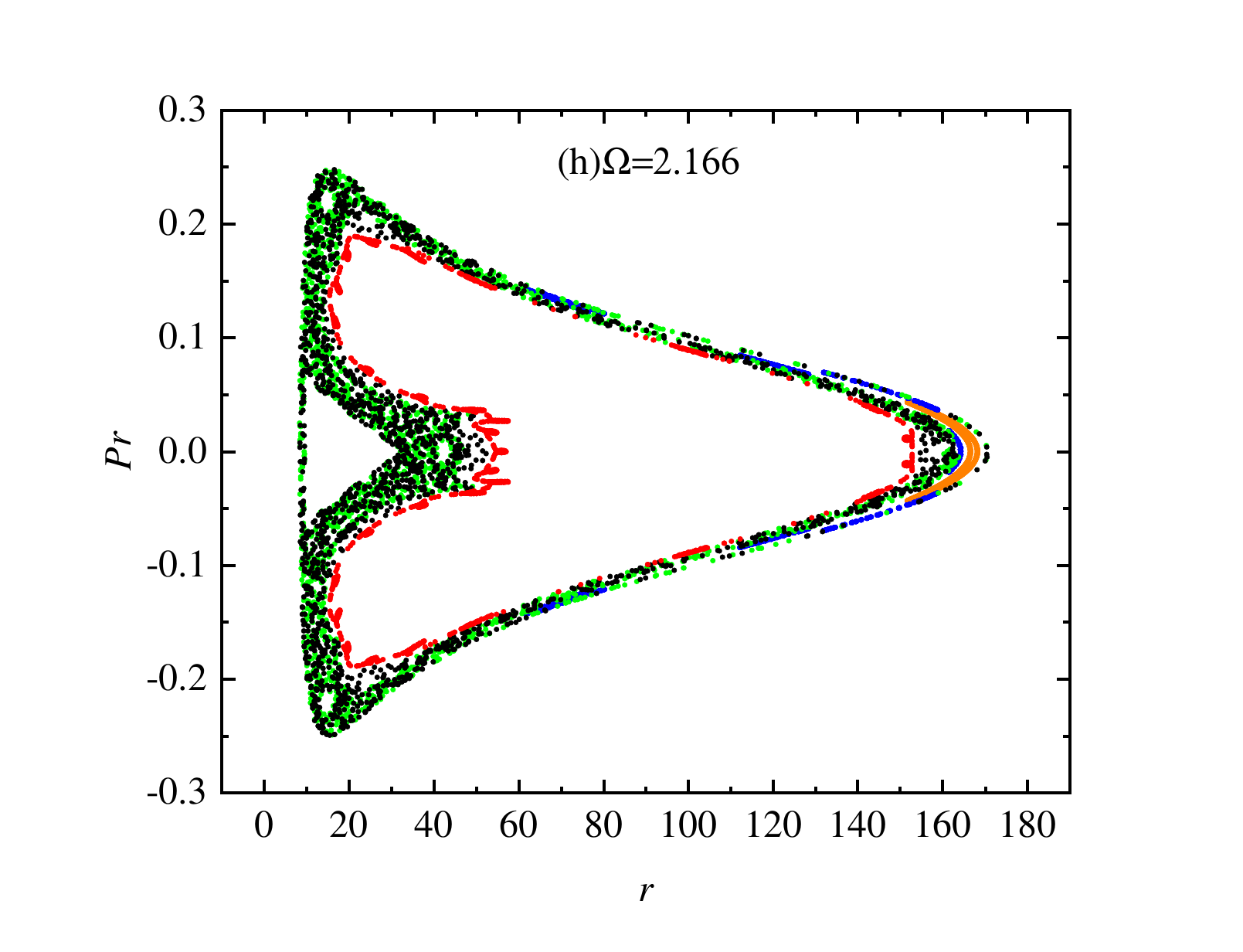}
        \includegraphics[width=13pc]{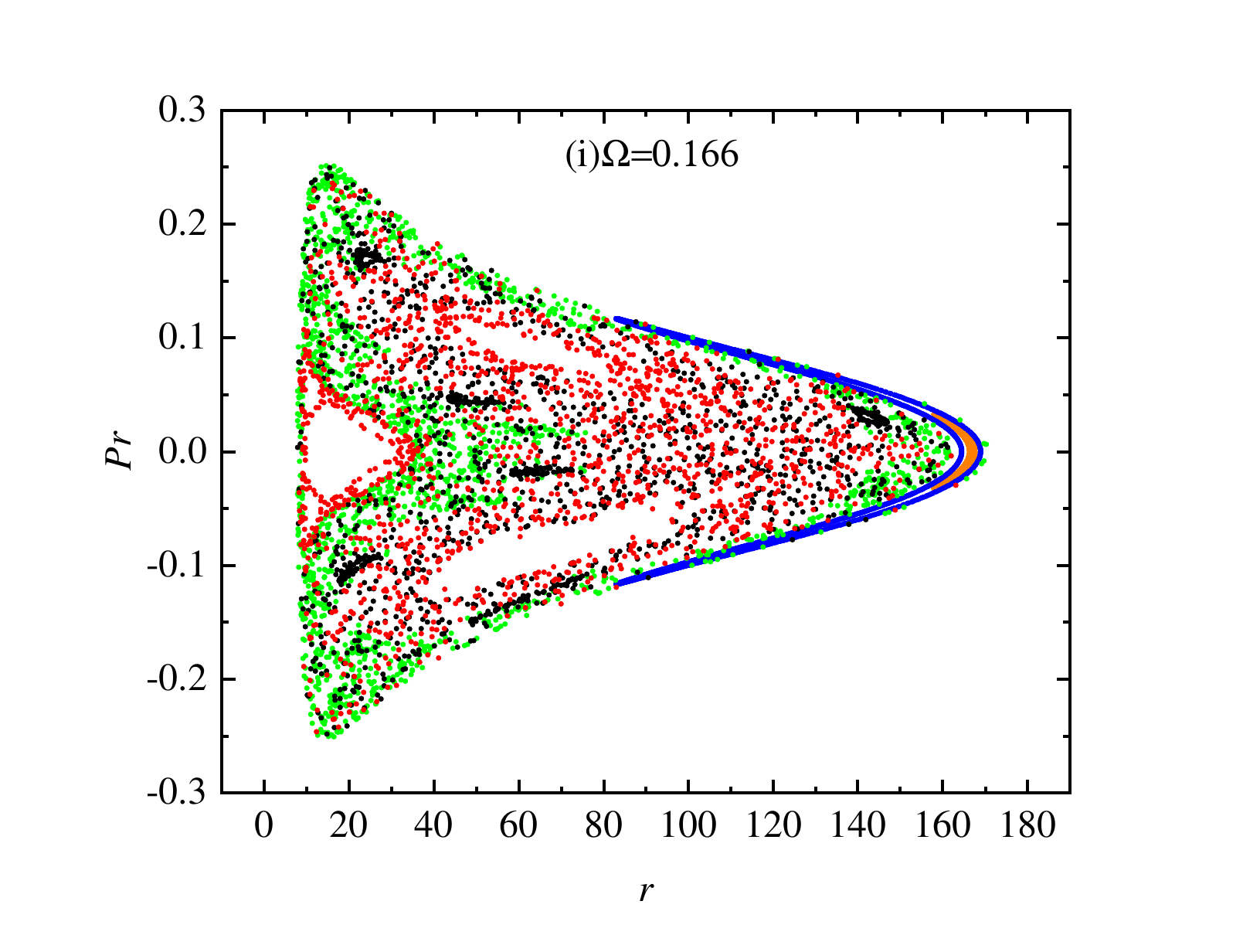}
        \caption{Poincar\'{e} sections.
      The parameters are $L=4.5$, $\gamma=0.089$, $\Omega=3.166$ and $\beta=8 \times 10^{-4}$, but three different values of the energy are (a) $E=0.999$, (b) $E=0.996$, and (c) $E=0.991$.
    The parameters are $\Omega=1.166$, $E=0.995$, $\gamma=0.089$, and $\beta=8 \times
    10^{-4}$,
    but three different values of the angular momentum are  (d) $L=6.5$, (e) $L=5.5$, and (f) $L=4$.
   The parameters are $\beta=8 \times 10^{-4}$, $E=0.995$, $L=4.5$, and $\gamma=0.089$,
   but three different values of $\Omega$ are
   (g) $\Omega=9.166$,  (h) $\Omega=2.166$, and  (i) $\Omega=0.166$.   }
    }
\end{figure*}

\begin{figure*}[htpb]
        \centering{
        \includegraphics[width=16pc]{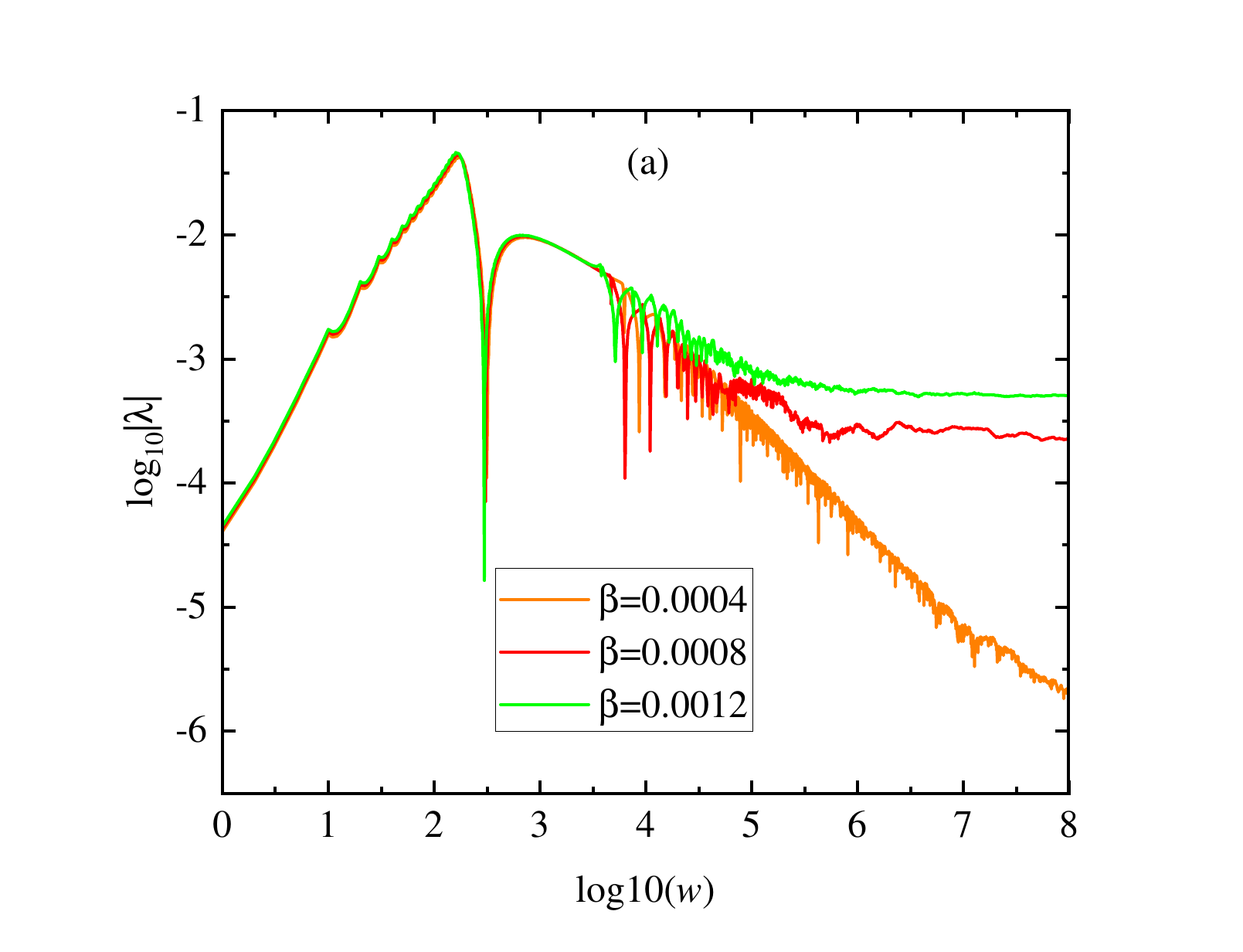}
        \includegraphics[width=16pc]{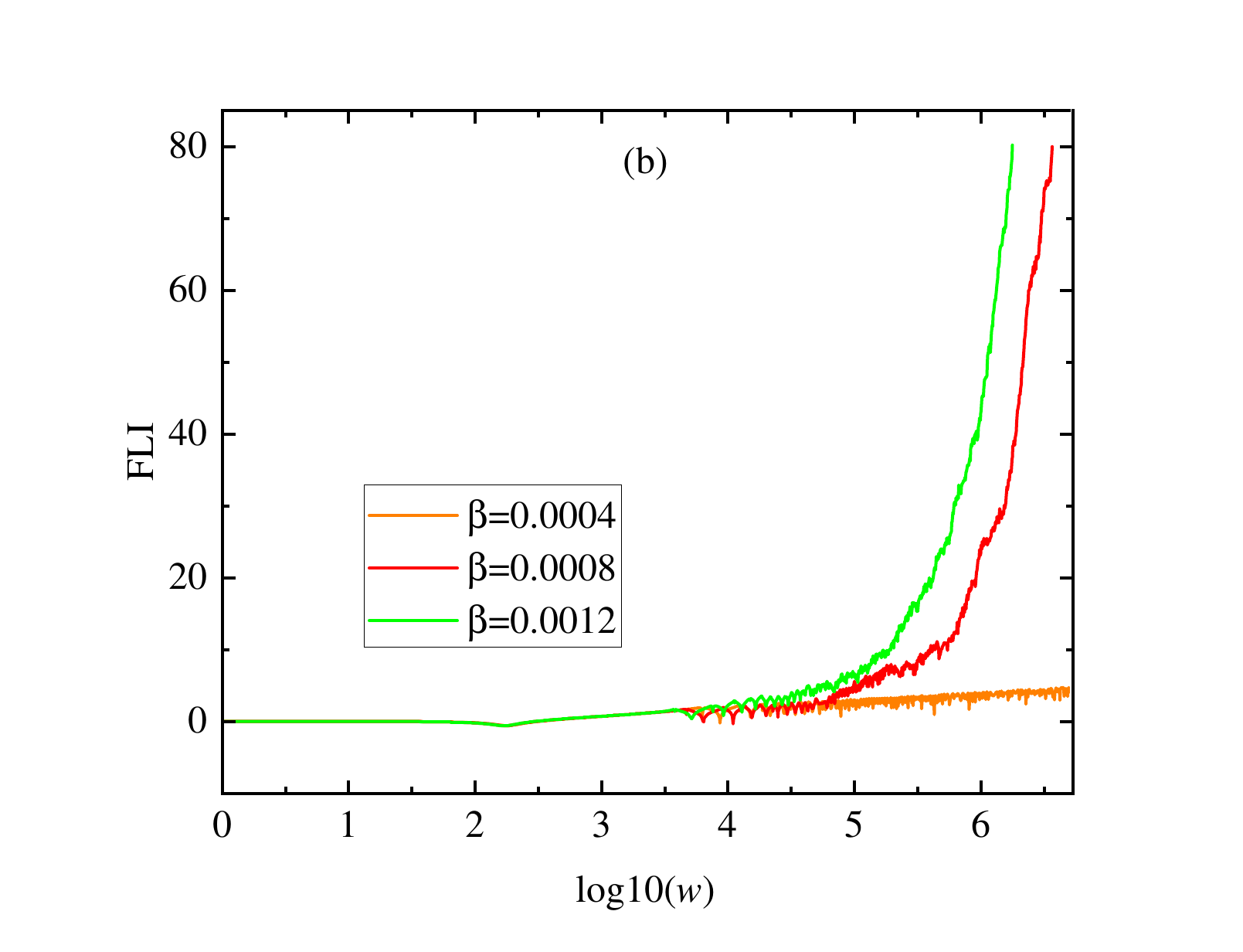}
        \includegraphics[width=16pc]{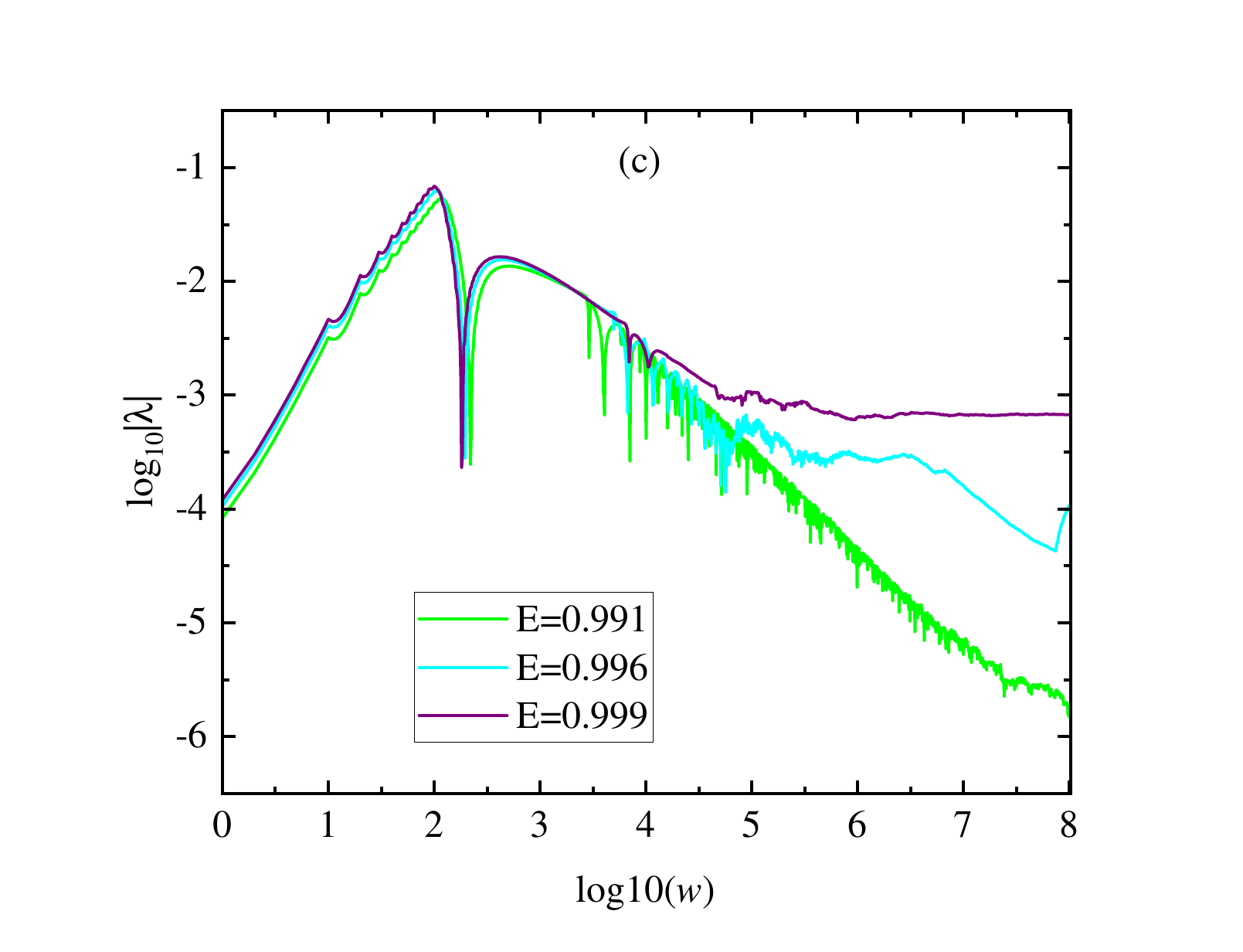}
        \includegraphics[width=16pc]{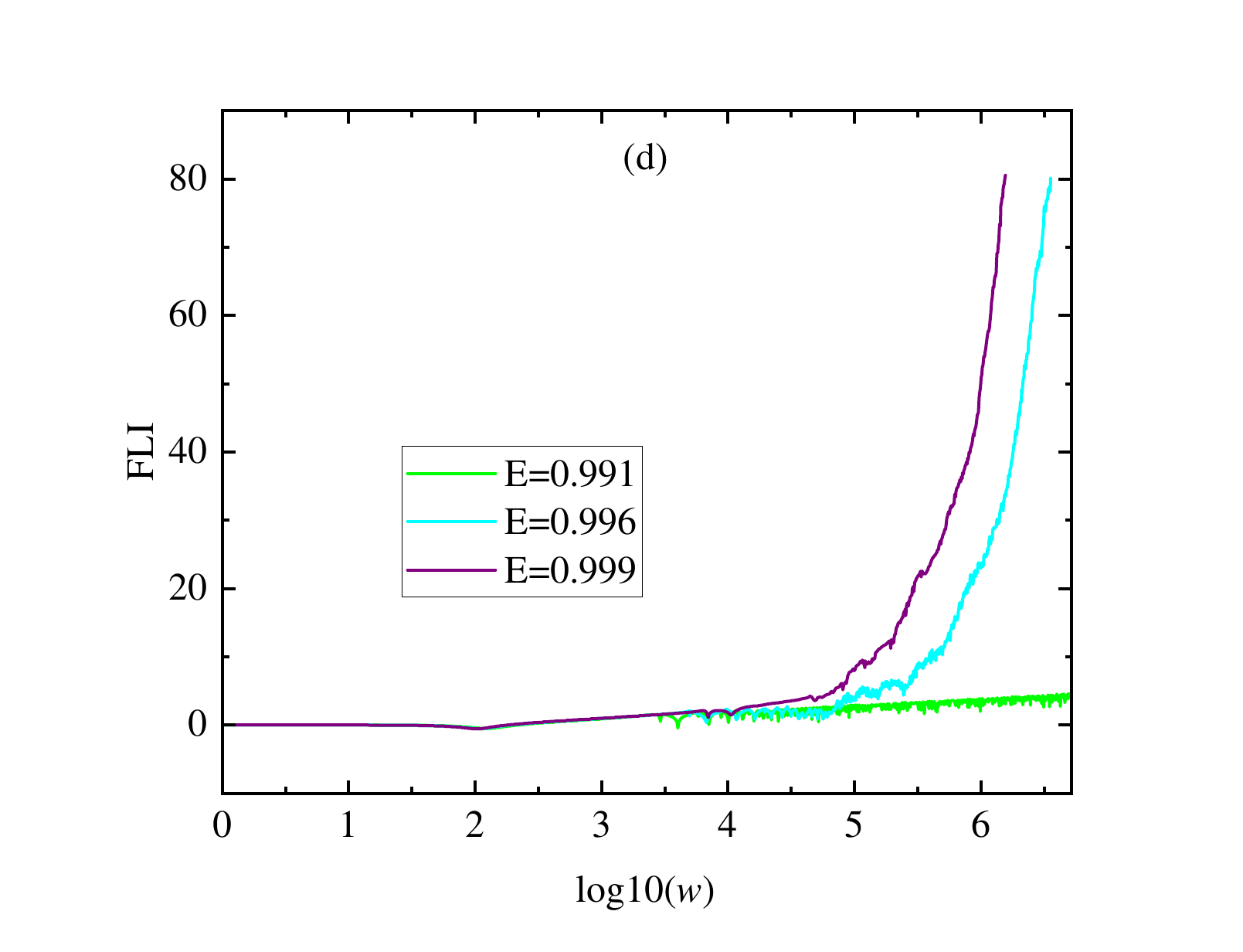}
        \includegraphics[width=16pc]{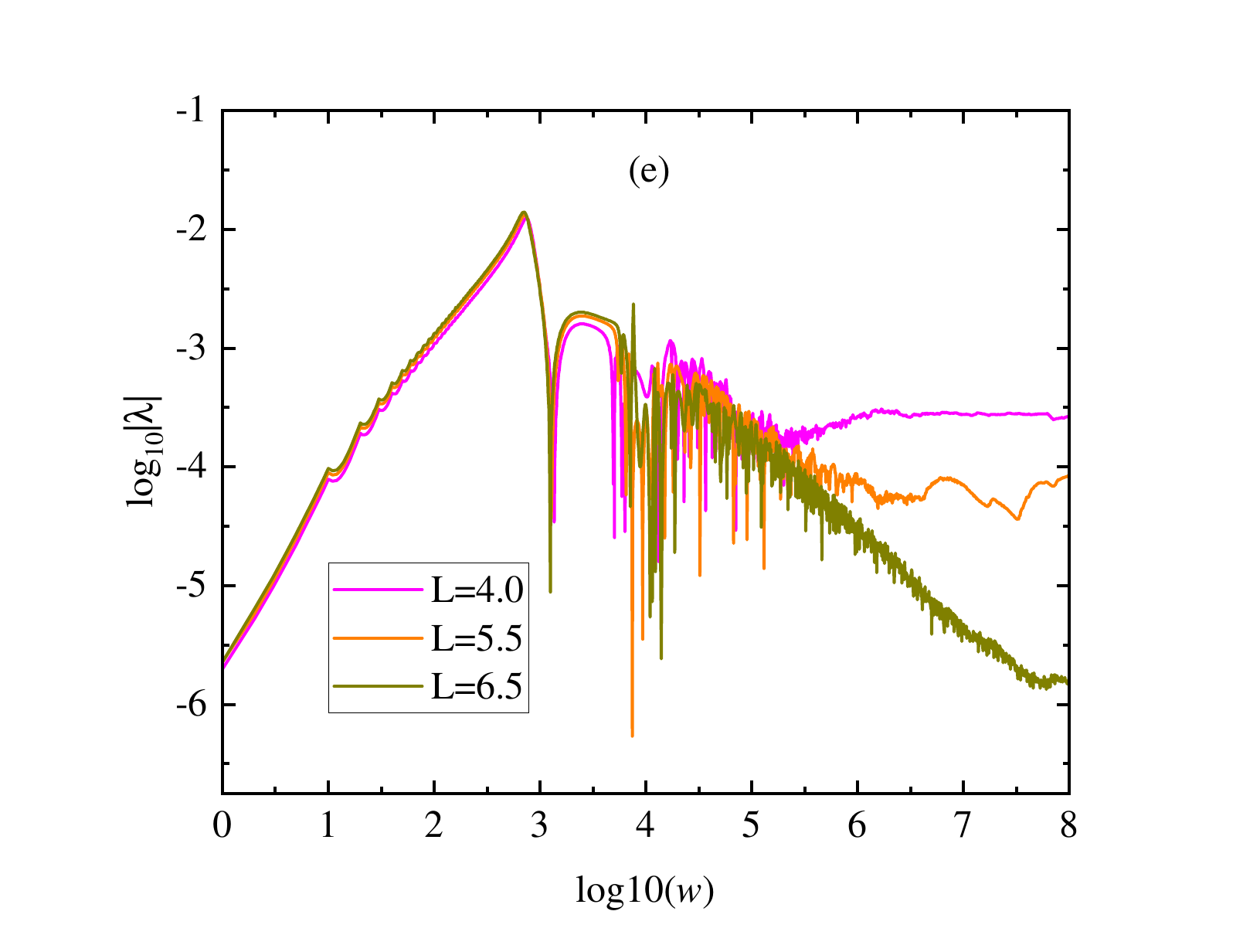}
        \includegraphics[width=16pc]{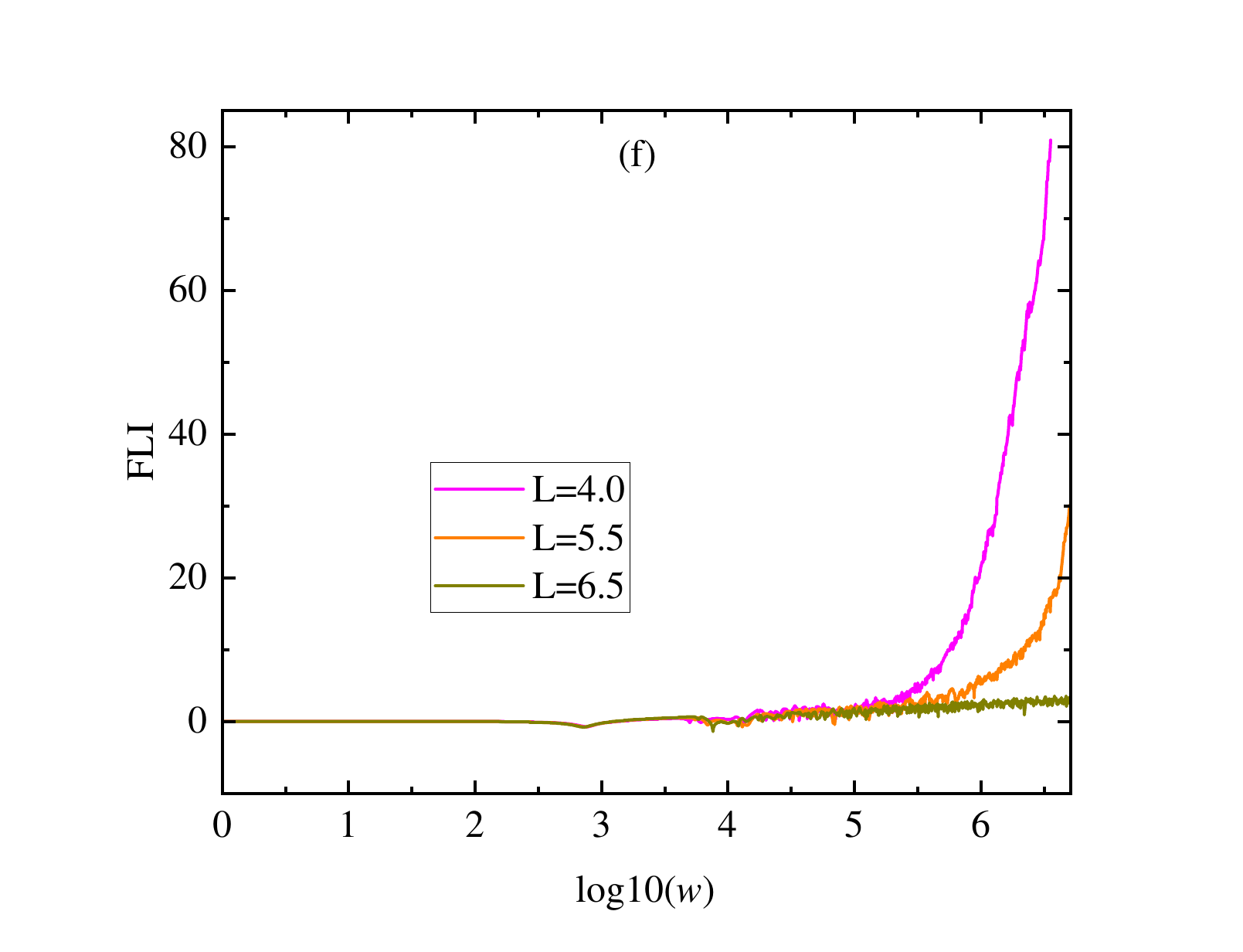}
        \includegraphics[width=16pc]{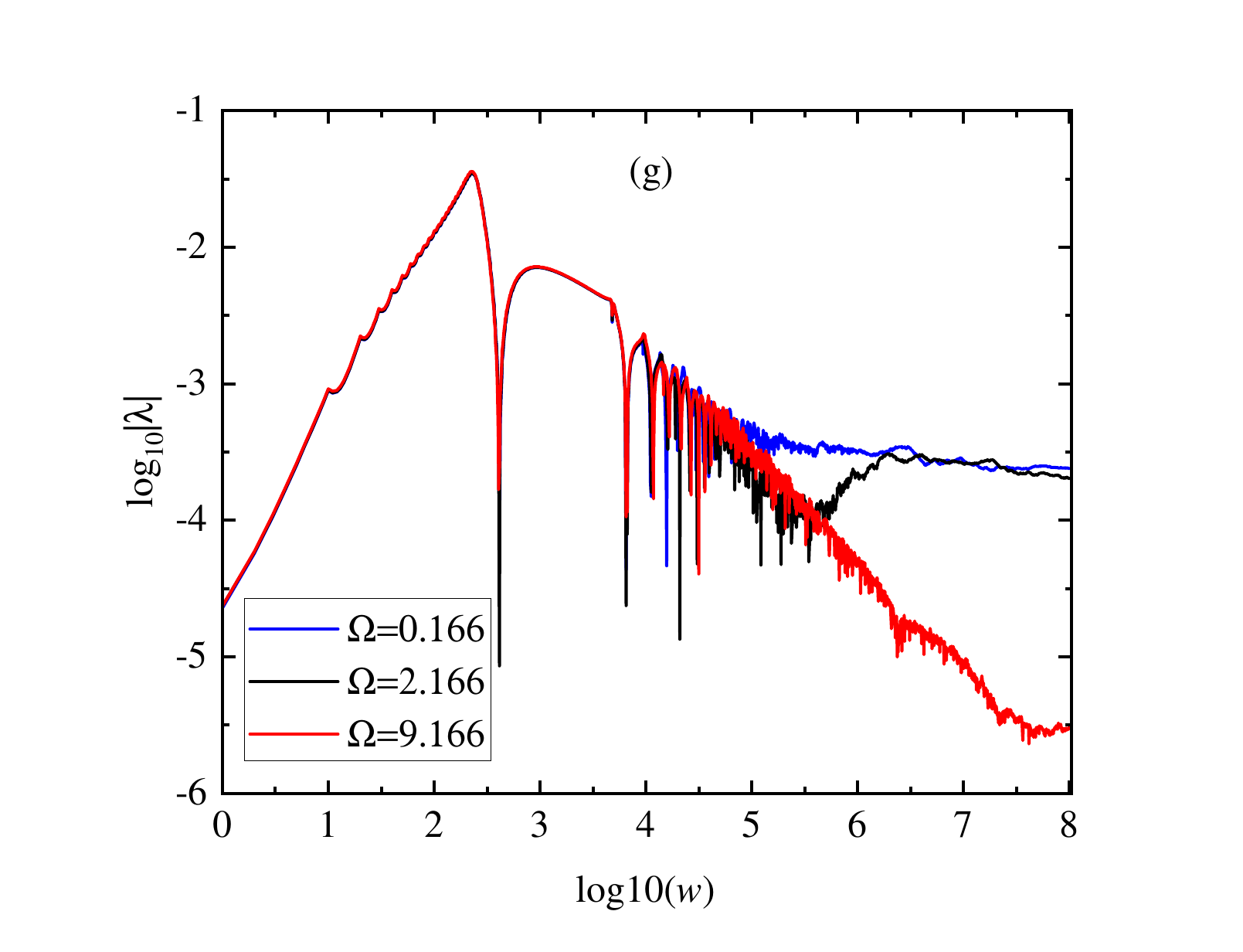}
        \includegraphics[width=16pc]{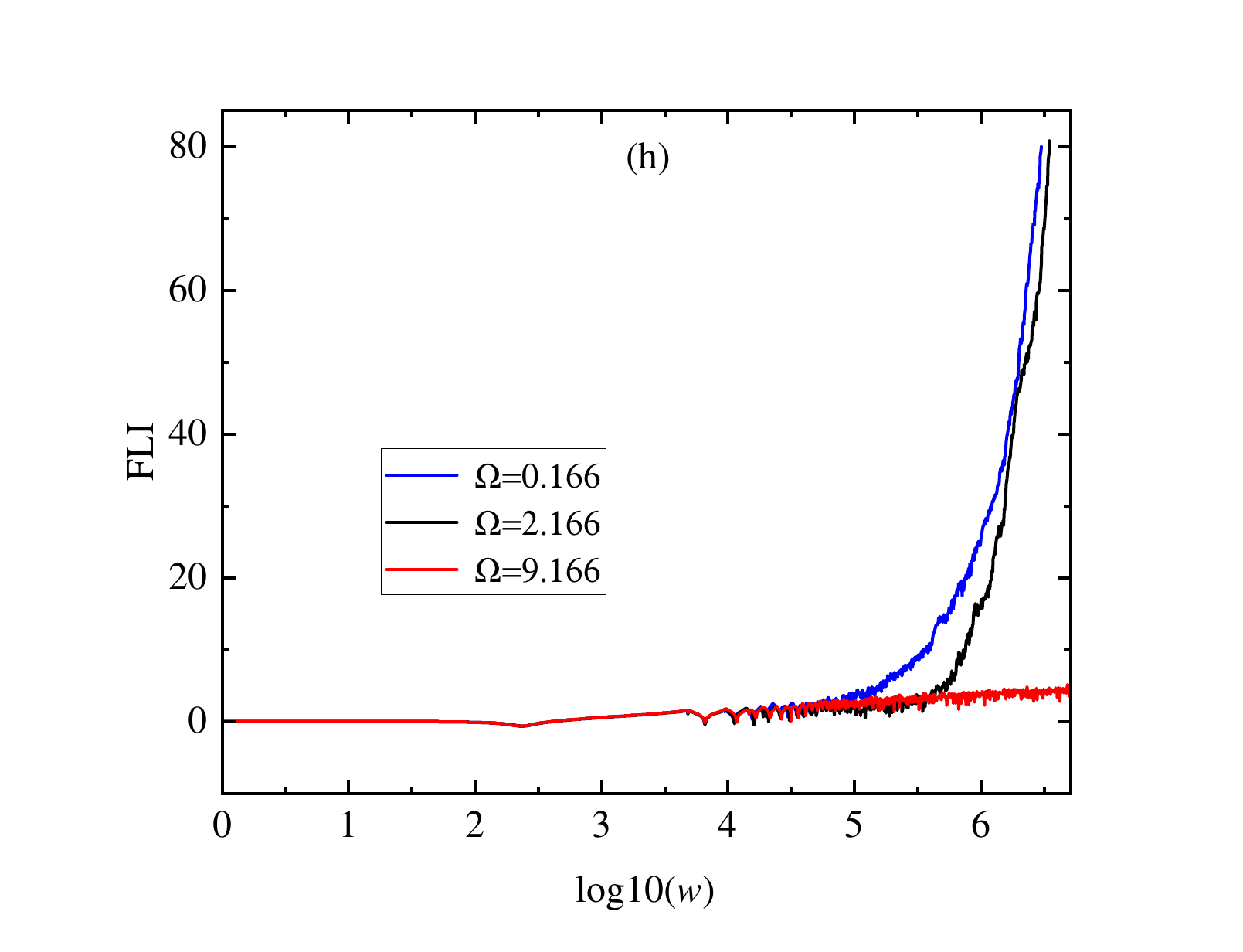}
        \caption{The maximal Lyapunov exponents $\lambda$ and the Fast Lyapunov Indicators (FLIs).
       (a) $\lambda$ and (b) FLIs for the tested orbit with the initial radius $r=33$,
  and the parameters $E=0.995$, $L=4.5$, $\gamma=0.089$ and $\Omega=2.166$, but three different values of the parameter $\beta$.
        (c)  $\lambda$ and (b) FLIs for the tested orbit having the initial radius $r=25$, and the parameters same as those of Figure 3a--c.
  (e)  $\lambda$ and (f) FLIs for  the tested orbit with the initial radius $r=80$ and the parameters corresponding to those of Figure 3d--f.
  (g)  $\lambda$  and (h) FLIs for the tested orbit having the initial radius $r=40$ and the parameters same as those of Figure 3g--i.
  }
    }
\end{figure*}

\begin{figure*}[htpb]
        \centering{
        \includegraphics[width=16pc]{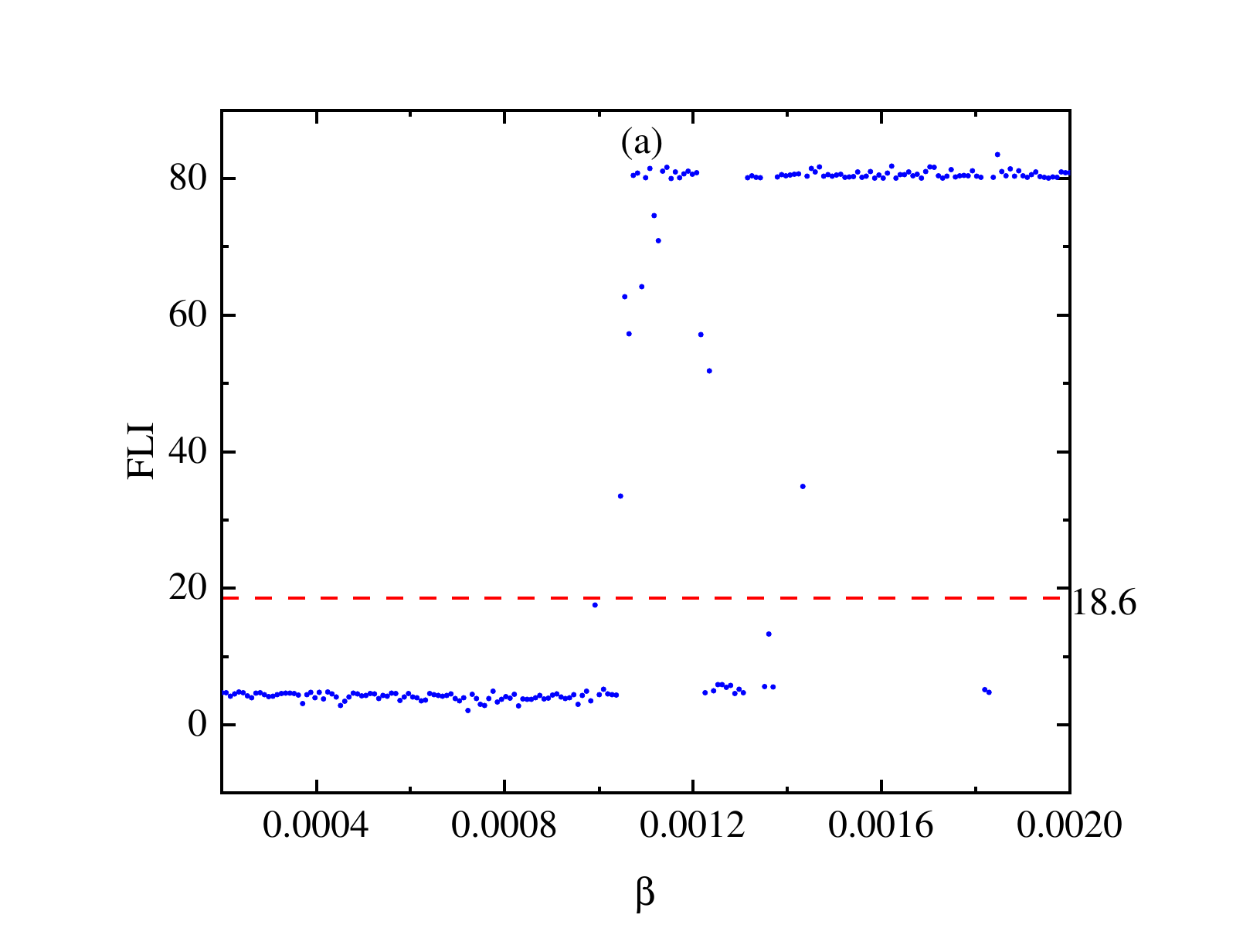}
        \includegraphics[width=16pc]{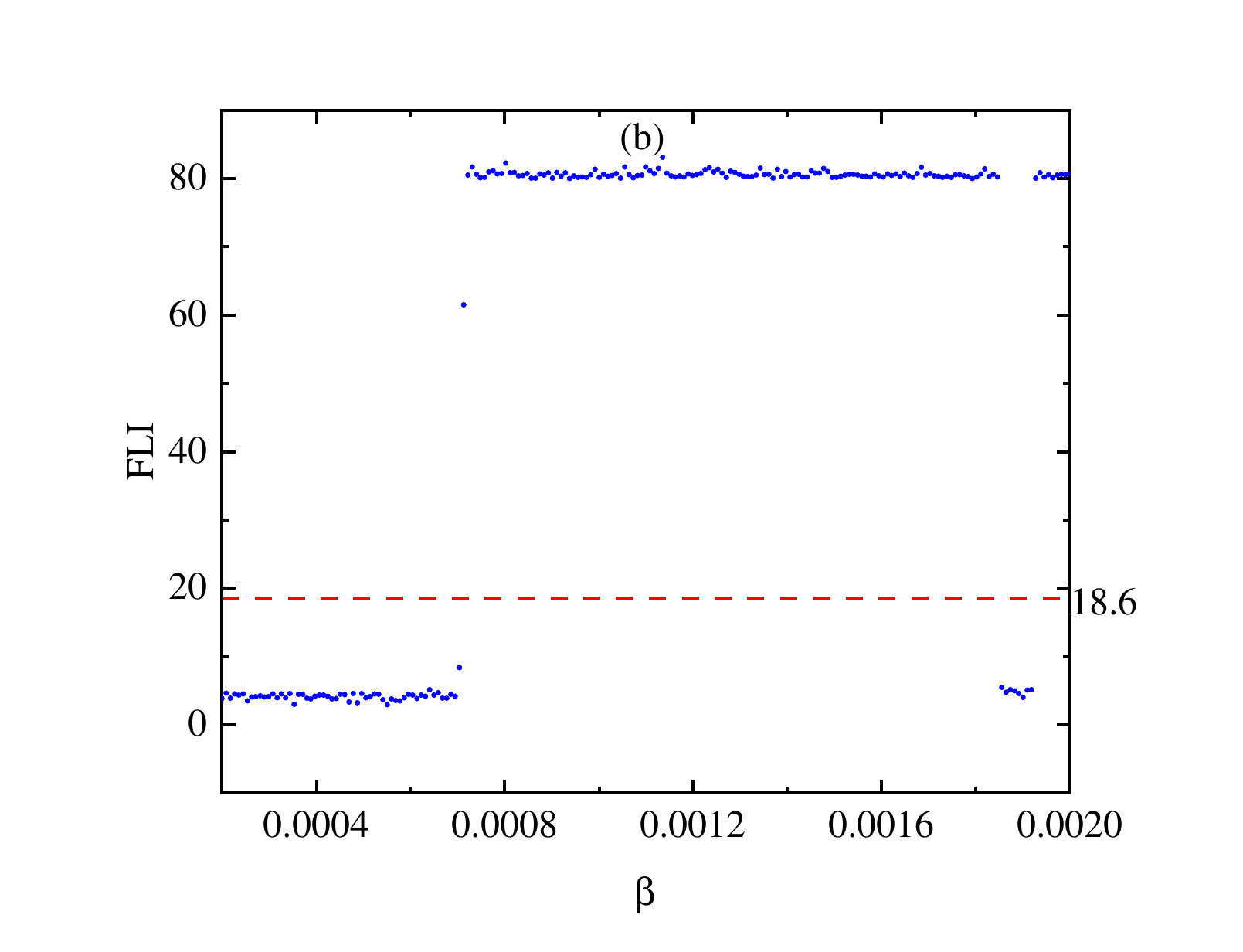}
 \caption{(a) Dependence of the FLI on the
magnetic parameter $\beta$ in Figure 2a-c. The initial radius is
$r=40$. The FLI for a given value $\beta$ is obtained after
integration time $w=5\times10^6$. The FLIs $\geq$ 18.6 indicate
disorder dynamics, while the FLIs $<$ 18.6 show regular dynamics.
(b) Dependence of the FLI on the magnetic parameter $\beta$ in
Figure 2d--f.}

    }
\end{figure*}

\begin{figure*}[htpb]
        \centering{
        \includegraphics[width=13pc]{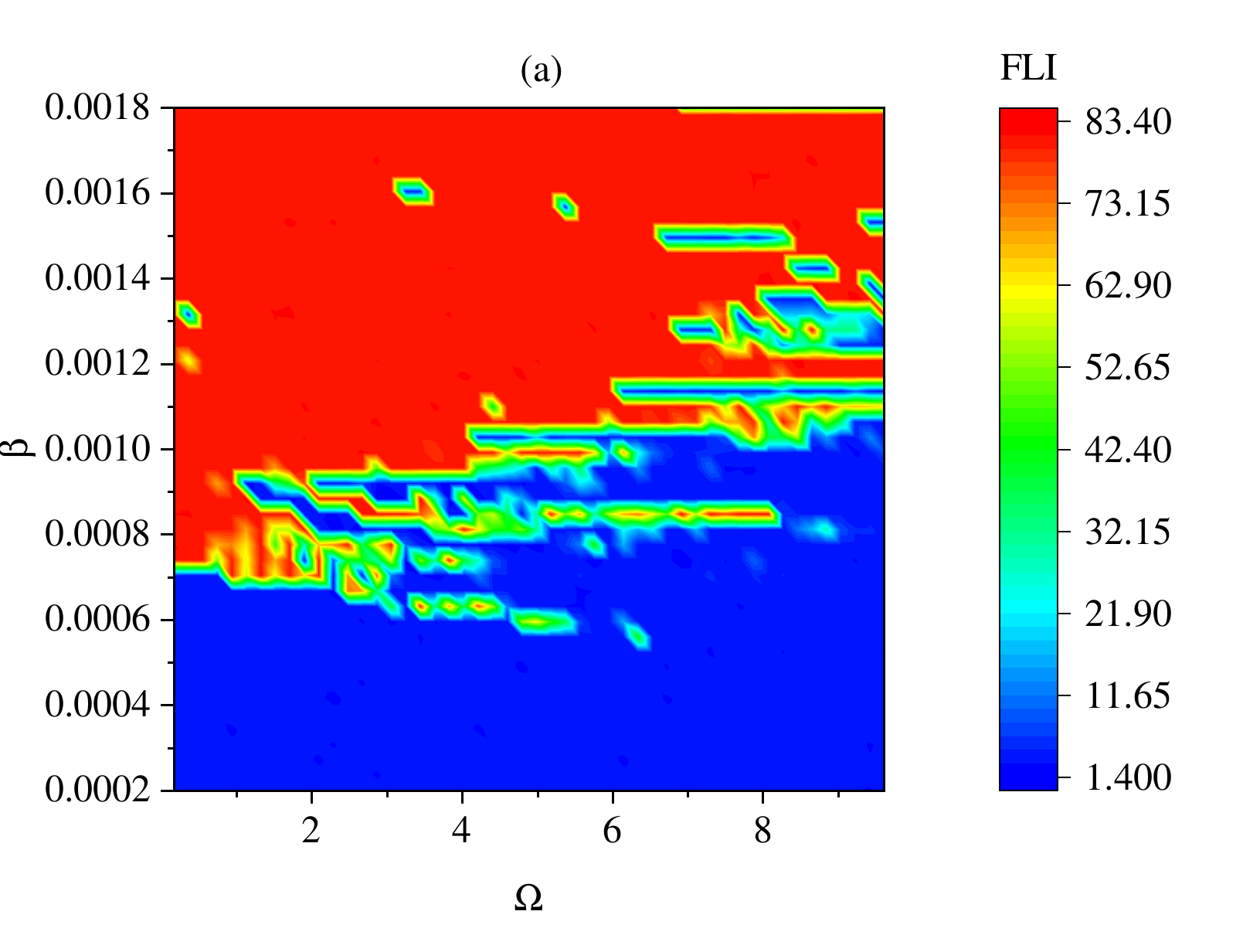}
        \includegraphics[width=13pc]{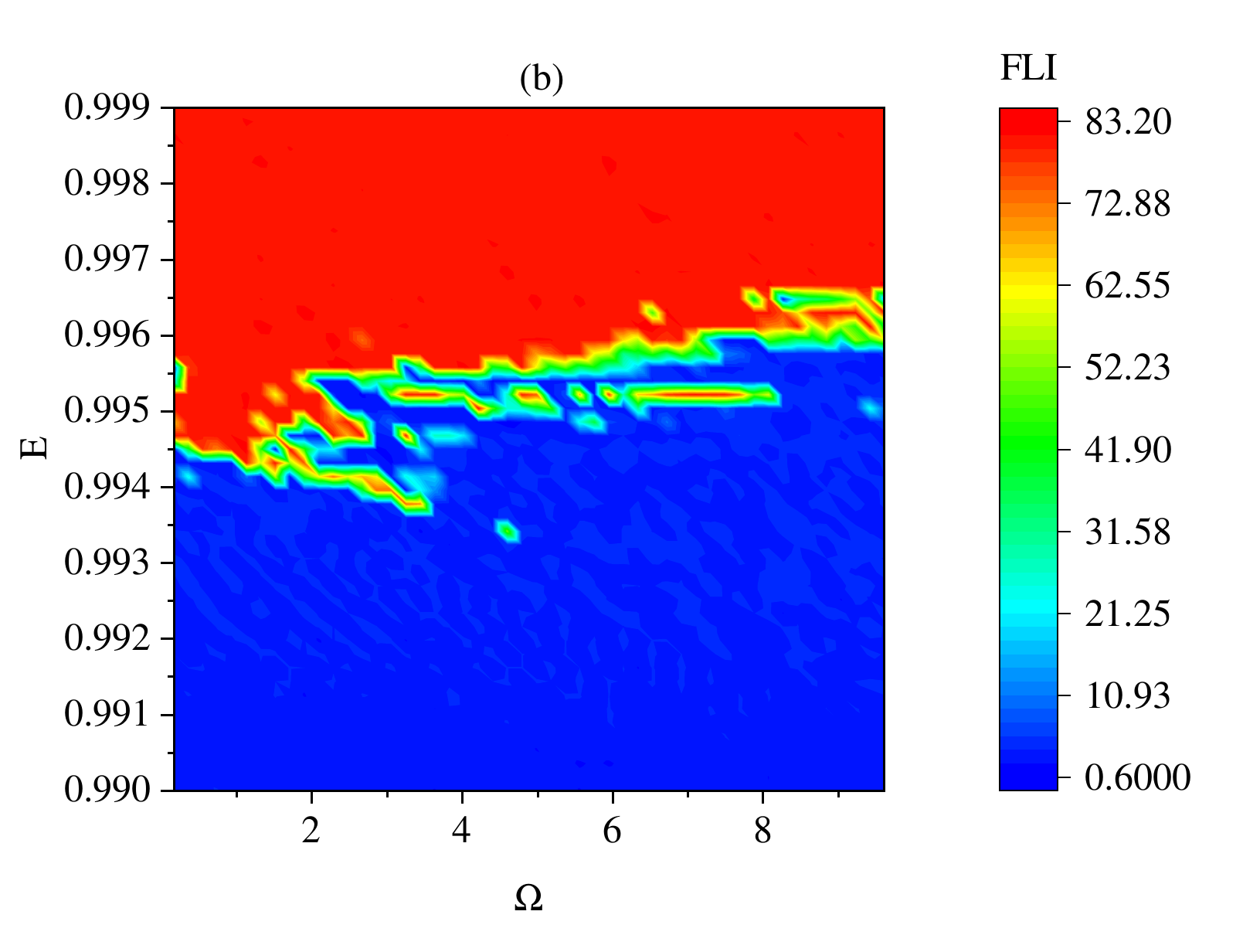}
        \includegraphics[width=13pc]{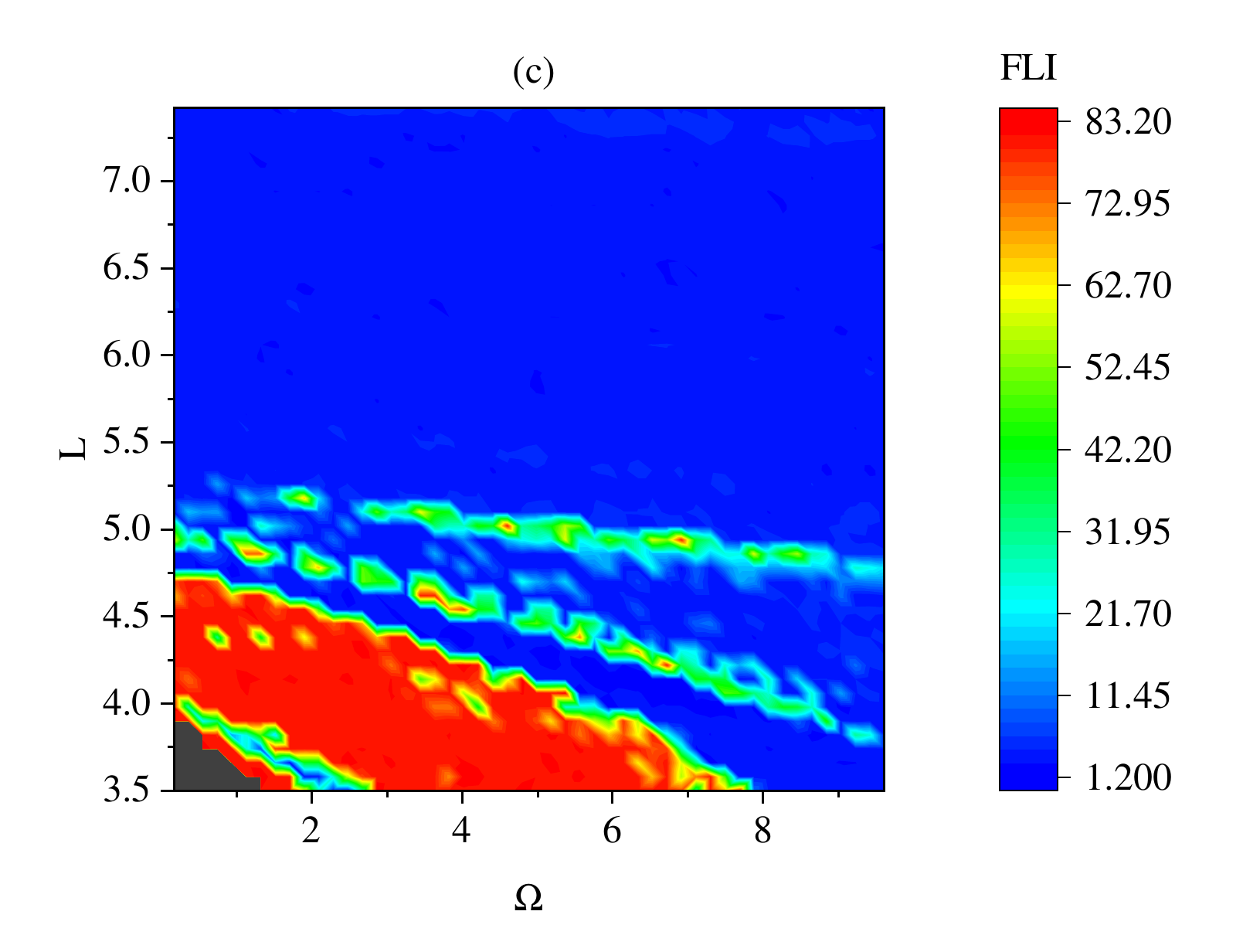}
\caption{Distributions of two parameters corresponding to order and chaos via the FLIs for the tested orbit with the  initial radius $r=40$.
(a) The two-parameter space is $(\Omega,\beta)$
 and the other parameters are $E=0.995$, $L=4.5$ and$\gamma=0.089$.
(b) The two-parameter space is $(\Omega,E)$ and the other parameters are the
same as those in Figure 3a.
(c) The two-parameter space is $(\Omega,L)$ and the other parameters are  the
same as those in Figure 3d.          }
    }
\end{figure*}

\begin{figure*}[htpb]
    \centering{
        \includegraphics[width=13pc]{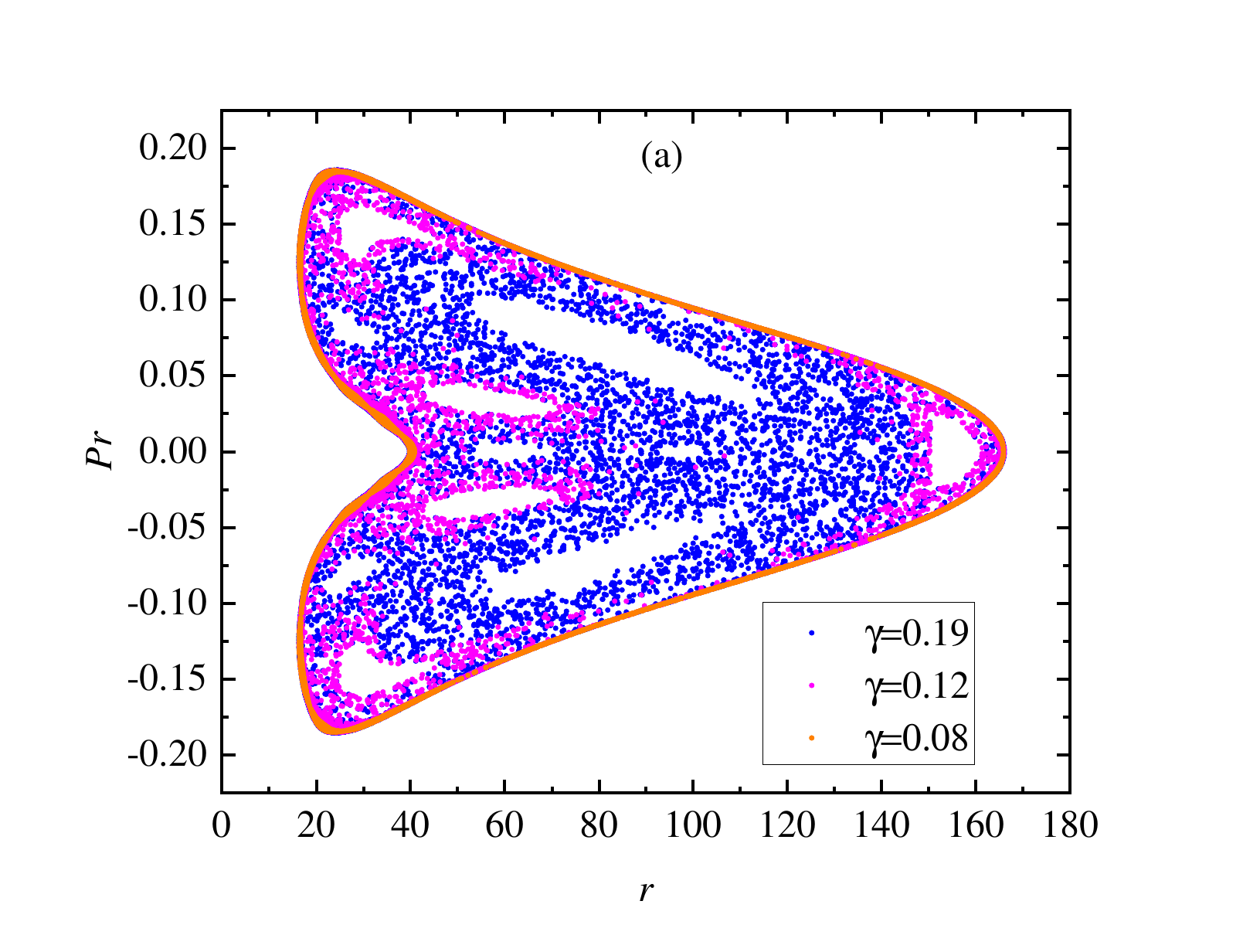}
        \includegraphics[width=13pc]{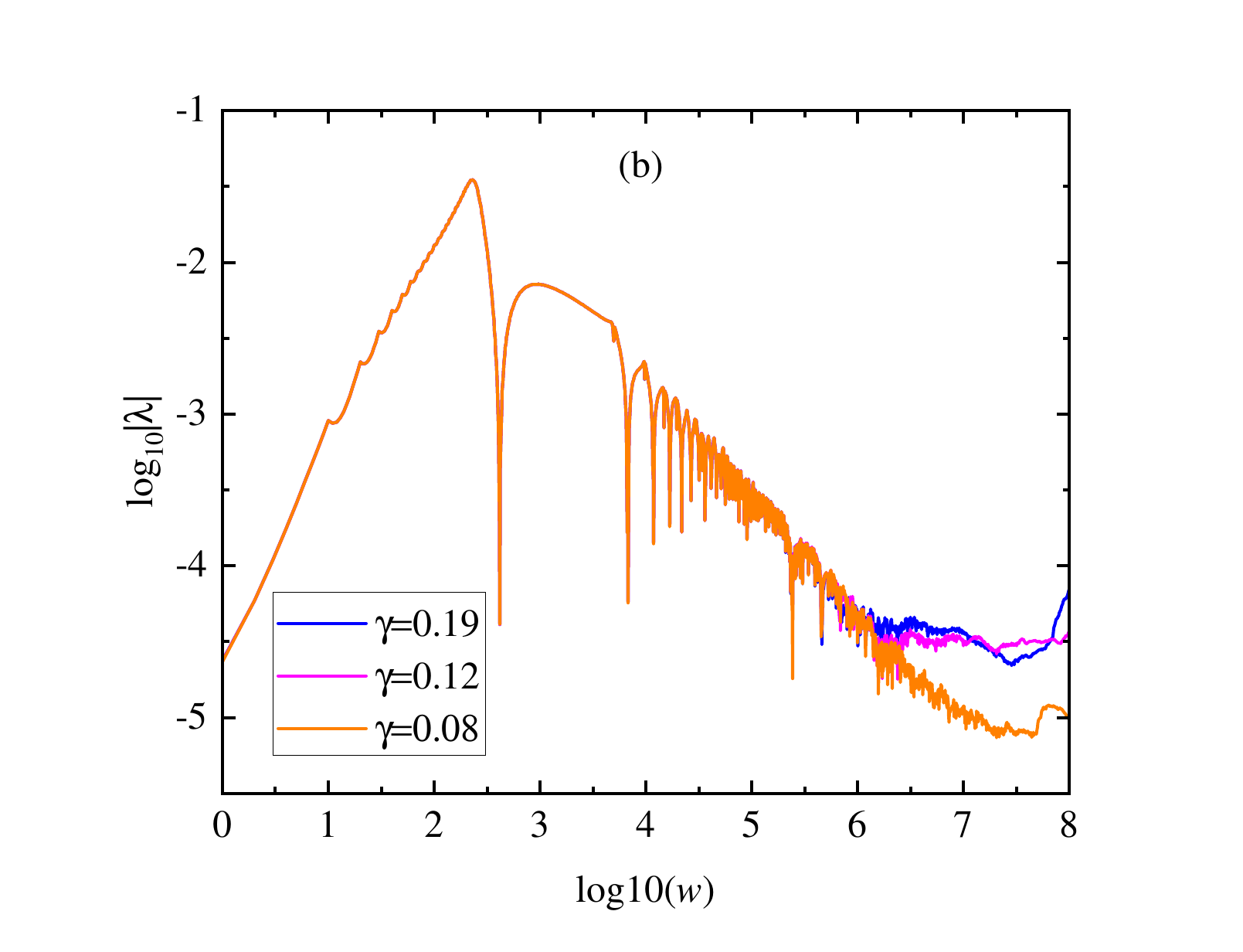}
        \includegraphics[width=13pc]{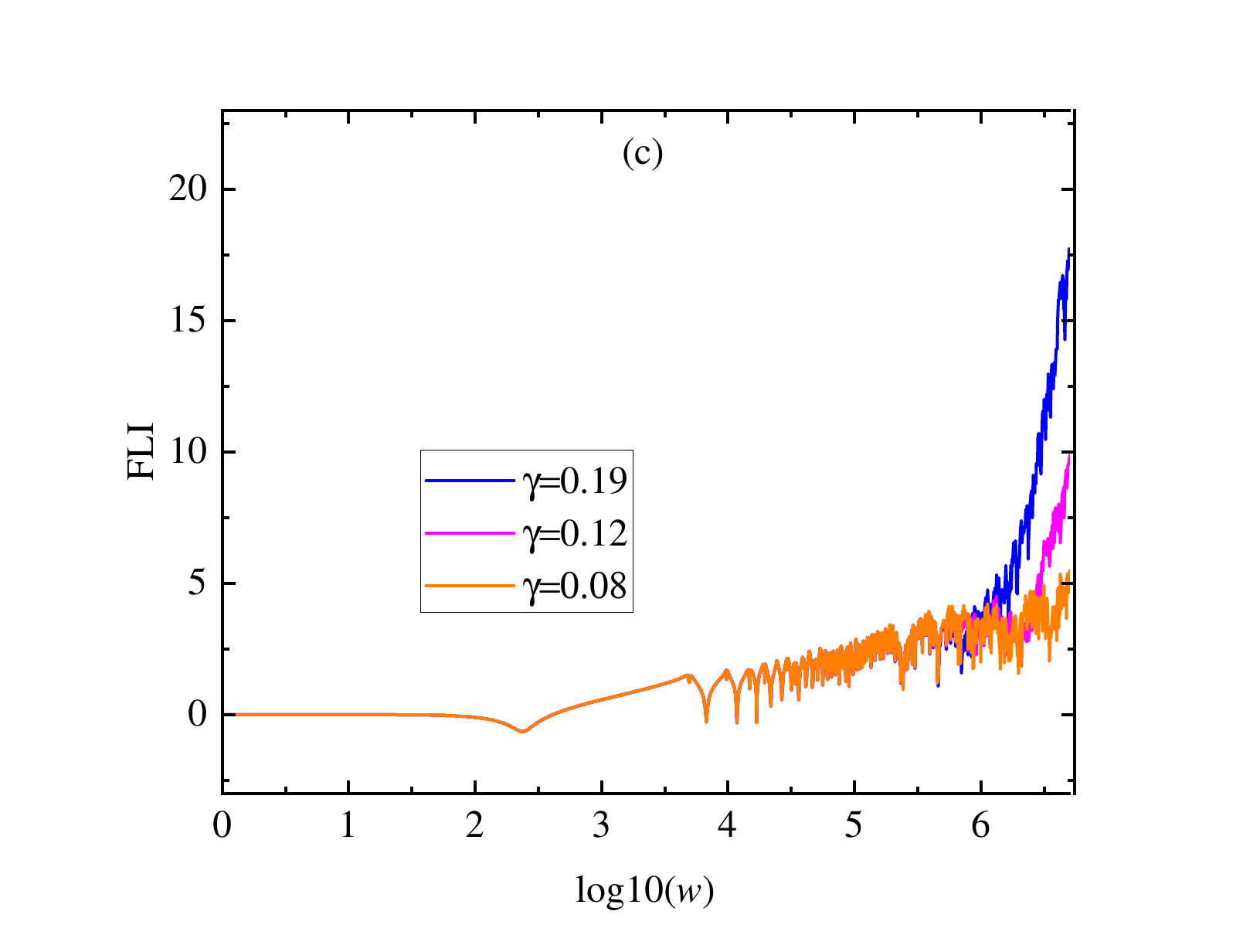}
        \includegraphics[width=13pc]{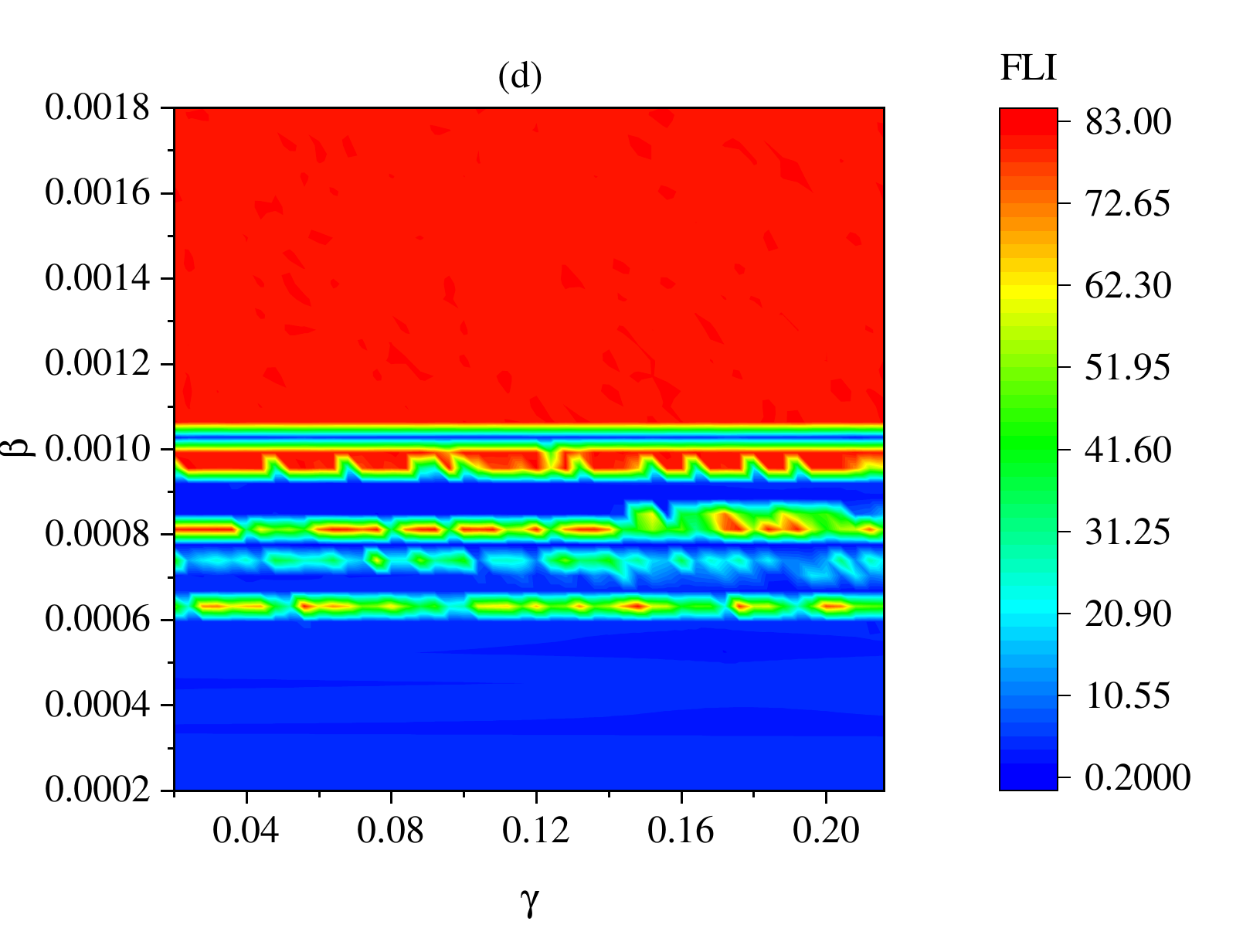}
        \includegraphics[width=13pc]{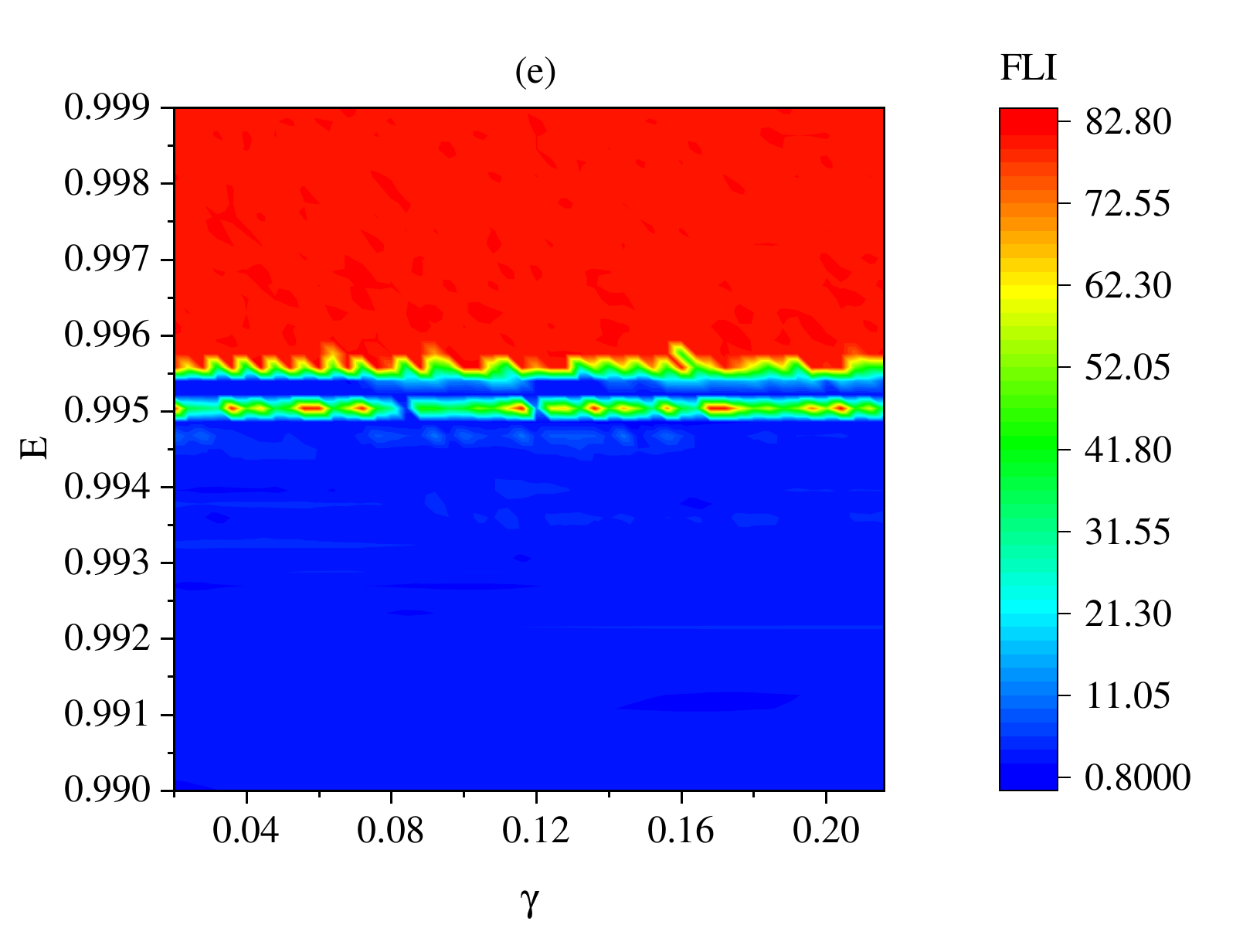}
        \includegraphics[width=13pc]{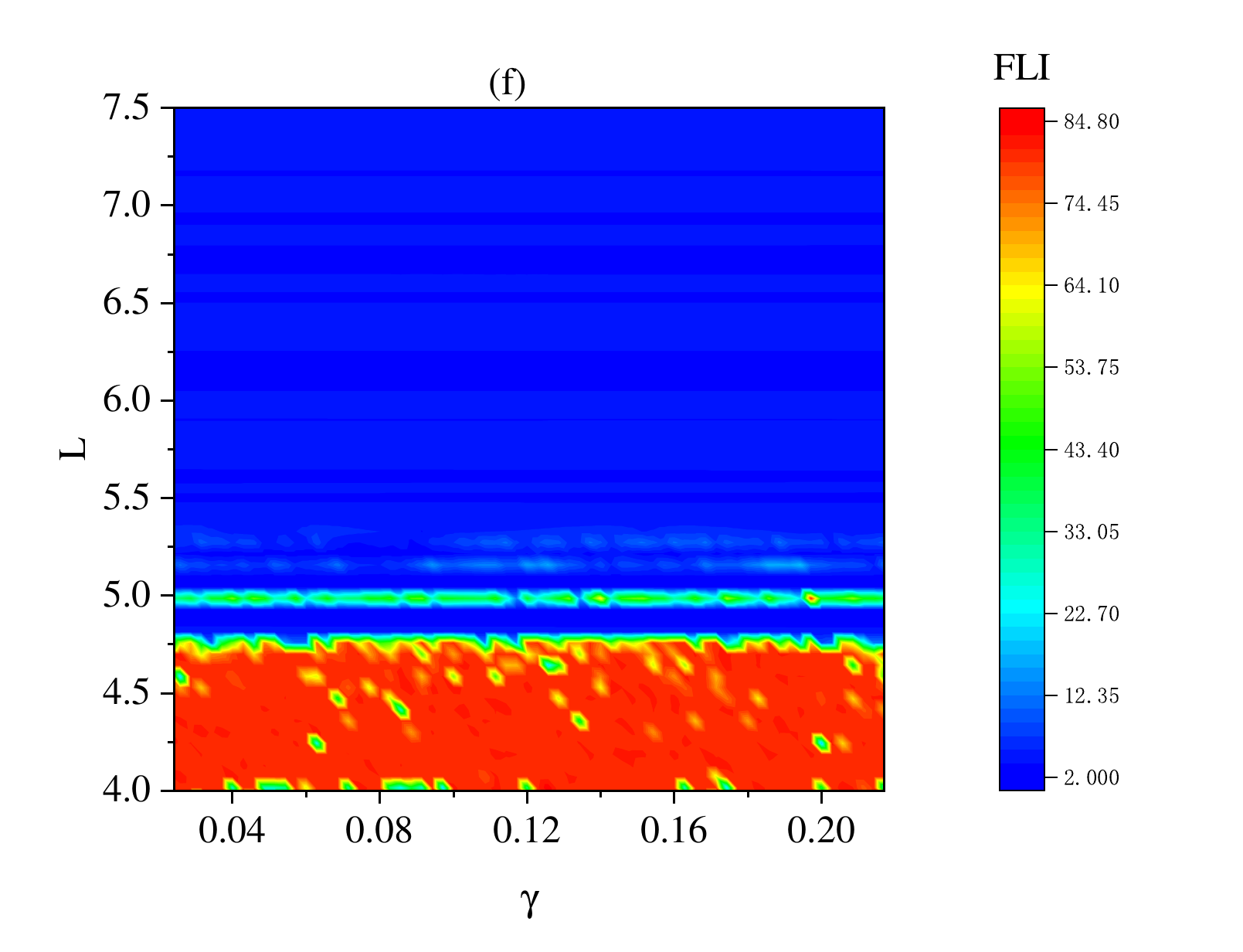}
\caption{The impact of the parameter $\gamma$ on the chaotic behavior.
(a) Poincar\'{e} sections, where the other parameters are $E=0.995$, $ L=5.157$, $\beta=8\times 10^{-4}$, $\Omega=0.166$ and $\gamma=0.08$ and
the tested orbit has its initial separation $r=40$.
(b) The maximal Lyapunov exponents $\lambda$ and (c) the FLIs for the three cases in panel (a).
(d) The FLIs for the two-parameter space $(\gamma,\beta)$, where the other
parameters are $E=0.995$, $\Omega=4.165 $, and $L=4.5$.
(e) The FLIs for the two-parameter space $(\gamma,E)$, where the other
parameters are $\beta=0.0008$, $\Omega=4.165 $, and $L=4.5$.
(f) The FLIs for the two-parameter space $(\gamma,L)$, where the other
parameters are $E=0.995$, $\beta=8\times 10^{-4}$, and $\Omega=0.166$.
          }
    }
\end{figure*}

\end{document}